\documentclass[a4paper,12pt]{article}
\usepackage{feynmp-auto,expdlist}
\usepackage{amsmath, amsfonts}
\usepackage{graphicx,arydshln}
\usepackage{enumerate}
\usepackage{hyperref}
\usepackage{latexsym}
\usepackage{dsfont}
\usepackage{hepnicenames}
\usepackage{enumerate}
\usepackage{soul}
\usepackage[normalem]{ulem}
\usepackage{comment}

\newcommand{\footnoten}[1]{}

\usepackage[font={small},textfont={it}]{caption} 

\usepackage{units}

\usepackage{mathrsfs,graphicx,rotating,amsmath,amsfonts,mathtools,booktabs,amssymb,wasysym}
\usepackage{hyperref}\usepackage{slashed}
\usepackage[nosort]{cite}
\usepackage[table,xcdraw,dvipsnames]{xcolor}
\usepackage{bm}
\usepackage{multirow,multicol}
\hypersetup{colorlinks,bookmarksopen,bookmarksnumbered,
	linkcolor=blus,pdfstartview=FitH,urlcolor=rossos,citecolor=verde}
\allowdisplaybreaks

\newcommand{\myfootnote}[1]{}
\newcommand{\myomit}[1]{{\color{gray}#1}}
\renewcommand{\myomit}[1]{}

\renewcommand{\[}{\left[}

\def\Lag{\mathscr{L}}

\newcommand{\mio}[1]{}

\newcommand{\med}[1]{\langle #1\rangle}

\def\bpm{\begin{pmatrix}}
	\def\epm{\end{pmatrix}}

\usepackage{mathrsfs}

\newcommand{\fig}[1]{~\ref{fig:#1}}

\renewcommand{\Im}{{\rm Im}\,}
\allowdisplaybreaks
\usepackage{multicol}
\usepackage{color}
\definecolor{rosso}{cmyk}{0,1,1,0.4}
\definecolor{rossos}{cmyk}{0,1,1,0.55}
\definecolor{rossoc}{cmyk}{0,1,1,0.2}
\definecolor{blu}{cmyk}{1,1,0,0.3}
\definecolor{blus}{cmyk}{1,1,0,0.6}
\definecolor{bluc}{cmyk}{1,1,0,0.1}
\definecolor{verde}{cmyk}{0.92,0,0.59,0.25}
\definecolor{verdec}{cmyk}{0.92,0,0.59,0.15}
\definecolor{verdes}{cmyk}{0.92,0,0.59,0.4}

\newcommand{\bp}{\bar{M}_{\rm Pl}}
\newcommand{\eq}[1]{~{\rm (\ref{eq:#1})}}

\newcommand{\GeV}{\,{\rm GeV}}
\newcommand{\TeV}{\,{\rm TeV}}

\def\circa#1{\,\raise.3ex\hbox{$#1$\kern-.75em\lower1ex\hbox{$\sim$}}\,}

\newcommand{\beq}{\begin{equation}}
\newcommand{\eeq}{\end{equation}}

\newcommand{\bea}{\begin{eqnarray}}
\newcommand{\eea}{\end{eqnarray}}
\newcommand{\be}{\begin{equation}}
\newcommand{\ee}{\end{equation}}
\font\tenrsfs=rsfs10 at 12pt
\font\sevenrsfs=rsfs7
\font\fiversfs=rsfs5
\newfam\rsfsfam
\textfont\rsfsfam=\tenrsfs
\scriptfont\rsfsfam=\sevenrsfs
\scriptscriptfont\rsfsfam=\fiversfs

\newsavebox\MBox

\renewenvironment{thebibliography}[1]
{\begin{multicols}{2}[\section*{\refname}]%
		\@mkboth{\MakeUppercase\refname}{\MakeUppercase\refname}%
		\list{\@biblabel{\@arabic\c@enumiv}}%
		{\settowidth\labelwidth{\@biblabel{#1}}%
			\leftmargin\labelwidth
			\advance\leftmargin\labelsep
			\@openbib@code
			\usecounter{enumiv}%
			\let\p@enumiv\@empty
			\renewcommand\theenumiv{\@arabic\c@enumiv}}%
		\sloppy
		\clubpenalty4000
		\@clubpenalty \clubpenalty
		\widowpenalty4000%
		\sfcode`\.\@m}
	{\def\@noitemerr
		{\@latex@warning{Empty `thebibliography' environment}}%
		\endlist\end{multicols}}

\newcommand{\eV}{\,{\rm eV}}
\newcommand{\SU}{\,{\rm SU}}

\renewcommand{\L}\Lag

\def\circa#1{\,\raise.3ex\hbox{$#1$\kern-.75em\lower1ex\hbox{$\sim$}}\,}
\makeatletter

\font\ital=cmu10

\def\hhref#1{\href{http://arxiv.org/abs/#1}{arXiv:#1}}
\usepackage{xstring}
\newcommand{\hhrefq}[1]{\IfSubStr{#1}{:}{\href{http://inspirehep.net/search?ln=en&ln=en&p=#1&of=hb&action_search=Search&sf=&so=d&rm=&rg=25&sc=0}{InSpire:#1}}{\hhref{#1}}}

\def\art{\@ifnextchar[{\eart}{\oart}}
\def\eart[#1]#2#3#4#5#6{{\rm #2}, {\em #3 \bf #4} {\rm (#6) #5} ({\em #1})}
\def\article{\@ifnextchar[{\earticle}{\oarticle}}
\def\oarticle#1#2#3#4#5#6{{\rm #1}, {\ital `#6'}, {\rm #2 #3 (#5) #4}}
\def\earticle[#1]#2#3#4#5#6#7{{\rm #2}, {\ital `#7'}, {\rm #3 #4 (#6) #5}  [\hhrefq{#1}]}
\def\hepart[#1]#2{{\rm #2, \sl#1}}
\def\heparticle[#1]#2#3{#2, {\ital `#3'} [\hhrefq{#1}]}

\newcommand{\hhrefqq}[1]{\IfBeginWith{#1}{10.}{\href{https://doi.org/#1}{doi:#1}}{\hhrefq{#1}}}
\def\earticle[#1]#2#3#4#5#6#7{{\rm #2}, {\ital `#7'}, {\rm #3 #4 (#6) #5}  [\hhrefqq{#1}]}

\renewenvironment{thebibliography}[1]
{\begin{multicols}{2}[\section*{\refname}]%
		\@mkboth{\MakeUppercase\refname}{\MakeUppercase\refname}%
		\list{\@biblabel{\@arabic\c@enumiv}}%
		{\settowidth\labelwidth{\@biblabel{#1}}%
			\leftmargin\labelwidth
			\advance\leftmargin\labelsep
			\@openbib@code
			\usecounter{enumiv}%
			\let\p@enumiv\@empty
			\renewcommand\theenumiv{\@arabic\c@enumiv}}%
		\sloppy
		\clubpenalty4000
		\@clubpenalty \clubpenalty
		\widowpenalty4000%
		\sfcode`\.\@m}
	{\def\@noitemerr
		{\@latex@warning{Empty `thebibliography' environment}}%
		\endlist\end{multicols}}

\newcounter{alphaequation}[equation]
\def\thealphaequation{\theequation\hbox to
	0.6em{\hfil\alph{alphaequation}\hfil}}
\def\eqnsystem#1{
	\def\@eqnnum{{\rm (\thealphaequation)}}
	\def\@@eqncr{\let\@tempa\relax \ifcase\@eqcnt \def\@tempa{& & &} \or
		\def\@tempa{& &}\or \def\@tempa{&}\fi\@tempa
		\if@eqnsw\@eqnnum\refstepcounter{alphaequation}\fi
		\global\@eqnswtrue\global\@eqcnt=0\cr}
	\refstepcounter{equation} \let\@currentlabel\theequation \def\@tempb{#1}
	\ifx\@tempb\empty\else\label{#1}\fi
	\refstepcounter{alphaequation}
	\let\@currentlabel\thealphaequation
	\global\@eqnswtrue\global\@eqcnt=0 \tabskip\@centering\let\\=\@eqncr
	$$\halign to \displaywidth\bgroup \@eqnsel\hskip\@centering
	$\displaystyle\tabskip\z@{##}$&\global\@eqcnt\@ne
	\hskip2\arraycolsep\hfil${##}$\hfil& \global\@eqcnt\tw@\hskip2\arraycolsep
	$\displaystyle\tabskip\z@{##}$\hfil
	\tabskip\@centering&\llap{##}\tabskip\z@\cr}
\def\endeqnsystem{\@@eqncr\egroup$$\global\@ignoretrue} \makeatother

\definecolor{Gray}{gray}{0.95}

\def\bal#1\eal{\begin{align}#1\end{align}}

\newcommand{\bAk}[3]{\langle #1  |  #2|#3  \rangle}

\def\citeall{\cite{Watkins:1991zt,1104.4793,1211.5615,1608.00583,2308.13070,2308.16224,2403.03252,2407.16747,2506.12123,2412.17912,2512.13815,2601.02458,2605.03758}}

\begin{document}
\thispagestyle{empty}
\begin{center}
{\LARGE \bf \color{rossos} Particle production from bubble collisions}\\[4ex]
{\bf\large Anish Ghoshal}$^a$, {\bf\large Pratyay Pal}$^b$, 
{\bf\large Alessandro Strumia}$^c$  \\[5mm]

{$^a$ \em Department of Physics and Astronomy, University of Sussex, UK}\\
{$^b$ \em Department of Physics, Oklahoma State University, USA}\\
{$^c$ \em Dipartimento di Fisica, Universit{\`a} di Pisa, Italia}\\[4ex]
{\bf\color{blus}\large Abstract}
\begin{quote}
  \color{blus}
Collisions of ultra-relativistic bubbles during cosmological phase transitions can 
produce particles much heavier than the transition scale.
Previous analyses modelled this process as the off-shell decay of the scalar background. 
We show that its results parametrically overestimate hard particle production and 
depend on the gauge choice and the coordinate choice in field space.
We propose an alternative formalism, analogous to the partonic description of high-energy collisions. In the ultra-relativistic limit, the colliding bubbles undergo nearly free passage and hard production arises from on-shell scatterings among the quanta constituting the Lorentz-contracted walls.
We apply this approach to heavy scalar, fermion, and vector particle production, and study the implications for dark matter, leptogenesis, graviton production and primordial gravitational waves.


\end{quote}
\end{center}

\setcounter{tocdepth}{1}
\tableofcontents

\newpage
\normalsize


\section{Introduction}

If a strongly first-order cosmological phase transition occurred in the early universe, 
it would have acted as a cosmic particle accelerator and collider. 
As vacuum-decay bubbles expanded, 
the released vacuum energy $\Delta V$ was largely converted 
into the kinetic energy of the bubble walls, whose Lorentz factor can become large at collision, $\gamma \gg 1$. 
Bubble collisions have therefore been proposed as non-thermal sources of heavy particles~\citeall. 
Although most of the energy released in the collision subsequently thermalizes, there are possible exceptions. 
In particular, if dark matter consists of a heavy stable particle, bubble collisions provide a significant non-thermal production mechanism~\cite{1211.5615,2403.03252,2605.03758}. 
The same collisions have been studied as sources of  baryogenesis~\cite{1104.4793,1608.00583},
right-handed neutrinos and leptogenesis~\cite{2407.16747,2601.02458}, gravitational waves~\cite{1608.00583,2412.17912}, cogenesis~\cite{2506.12123}.

However, computing heavy particle production from wall collisions is a non-trivial Quantum Field Theory problem.

In section~\ref{sec:bad} we summarise the Schwinger-like approach
assumed in previous literature~\citeall:
it interprets the field configuration of the scalar involved in the bubble collision
as off-shell quanta with density $f(p^2)$,
and convolves it with the imaginary part of the scalar off-shell propagator $\Im\Pi(p^2)$
to obtain their decay rate into heavy particles.
However, both factors are problematic.
The off-shell propagator yields gauge-dependent rates for vector production~\cite{2403.03252}.
We show that a more general problem affects production of scalars or fermions too: 
$\Pi(p^2)$  depends on the parameterisation chosen in field space.
As a consequence, the inferred production rate depends on arbitrary choices 
and can be non-vanishing when the underlying theory is free.
In general, off-shell propagators are unphysical.
Furthermore, the off-shell density $f(p^2)$ is computed under approximations such as perfectly elastic collisions
with the result that all quanta in the bubbles go far off-shell.
This parametrically over-estimates the hard production rate.
We clarify that the Schwinger-like approach is applicable to fields with negligible self-interactions,
which is not the case for bubble collisions.

\smallskip

In section~\ref{sec:good} we propose a Feynman-like simple approach 
that yields a gauge invariant rate, independent of field parameterisations. 
The Schwinger formalism was developed to compute $e^-e^+$ production in an electric field, 
where many quanta collectively accumulate large energy.
Instead, the wall of an ultra-relativistic bubble is a microscopically thin and Lorentz-contracted configuration.
The collision among two walls has such a short duration  that different pieces of each wall don't have time to interact.
The short-distance quanta that compose each wall therefore mostly see the short-distance modes of the other wall as quasi-free incoming particles; 
only those pairs that undergo an ordinary hard scattering produce heavy particles.
We thereby expand each ultra-relativistic colliding wave in terms of quasi-real `partonic' quanta.
The production rate of heavy particles is then approximated as `partonic' cross-sections.
The resulting rate of hard processes is parametrically different from the previous approximation.
In collider physics, this logic is known as the partonic approximation to the collision of two boosted hadrons.
Our claim is not that coherent wall dynamics is irrelevant.
Coherent evolution, rollback and re-collisions occur at later times and can be important for reheating and for particle production at scales of order $m_s$.
Our claim is more specific: because this evolution starts continuously from the free-passage configuration, it does not restore the unsuppressed hard timelike tail up to invariant masses of order $\gamma m_s$ assumed in the perfectly elastic or inelastic ansatz.

In section~\ref{sec:cross} we compute the basic cross sections for production of scalars, fermions, vectors.

In section~\ref{sec:models} we consider the consequences for dark matter.
We also compute specific models where dynamical symmetry breaking \`a la Coleman-Weinberg 
triggered by dark matter produces a first order phase transition. 

In section~\ref{sec:leptog} we consider the consequences for leptogenesis.

In section~\ref{sec:GW} we consider the consequences for gravitational waves and for gravitons
sourced by hard  particle production.

Conclusions are given in section~\ref{sec:concl}.

\section{The  off-shell approximation and its problems}\label{sec:bad}
We want to compute the production of particles with heavy mass $M$
from ultra-relativistic collisions of bubble walls during a first order cosmological phase transition driven by a real scalar $s$
with mass $m_s$.
We assume that the scalar potential $V(s)$ has two minima, and that
during the phase transition its vacuum expectation value changes by $\Delta s$
and the potential energy by $\Delta V$.

\subsection{Summary of the formalism}
The formalism of~\citeall{}
assumes that a classical scalar field configuration $s_0(x_\mu)$ describes
the two moving and colliding bubbles, their collision and their subsequent oscillations.
As usual, the imaginary part of the quantum effective action controls vacuum persistence as $e^{i\Gamma[s]}\sim \langle {\rm out}|{\rm in}\rangle_s$.
So  the probability $\wp$ of particle production at leading order is obtained as 
\beq\wp=2\Im\Gamma[s_0].\eeq
This is a correct but formal statement about the exact effective action evaluated on an exact solution. 
The practical approximation used in the literature makes two additional steps: 
it truncates the action at quadratic order in the wall field $s$ 
and evaluates the result on an approximate collision profile~\citeall. 
It is this combined approximation that we scrutinise below.

\smallskip

The action truncated at quadratic level in $s$ is written in terms of the $s$ quantum propagator $\Pi(p^2)$ as
\begin{equation}\label{eq:ImGamma}
    \text{Im} \Gamma[s_0]  \approx  \frac12 \int\frac{d^4 p}{(2\pi)^4} |s_0(p)|^2 \text{Im} \Pi(p^2)
\end{equation}
where the Fourier transform of the colliding field configuration $s_0(x)$ is denoted as $s_0(p)$.
It is significantly off-shell because $s_0(x)$ significantly deviates from a free wave, 
if $s$ self-interactions have a big effect during the collision.
This is the heart of the problem and it is what allows heavy particle production: the heavy off-shell quanta $s^*_p$ decay into pairs of particles $f$
with mass $2M < p$ giving rise to an imaginary part of $\Pi(p^2)$
as dictated by the optical theorem
\beq 
\Im\Pi (p^2) = \frac12 \sum_f \int d\Phi |\mathscr{A}(s^*_p \to f)|^2.\eeq
Specializing the above formalism to two walls moving along the $x$ axis, the number of particles produced per unit area $A$ in the  frontal collision is~\citeall
\beq \label{eq:frontaloff}
\frac{N}{A}
\approx \frac{1}{2\pi^2}\int  d\hat{s}\, f(\hat{s})  \Im \Pi(\hat{s}) \eeq
where $\hat s = p^2$ and $f(\hat{s})$ is related to the square of the Fourier transform of $s_0(x)$.
Approximating the collision as Perfectly Elastic one gets, according to~\citeall
\begin{equation}
f_{\rm PE}(\hat{s})
=
\frac{16 \, \Delta s^2}{\hat{s}^2} \ln
\frac{ 2 -\hat{s}\ell^2  + 2 \sqrt{1-\hat{s}\ell^2}}{  \hat{s}\ell^2 }\qquad\hbox{for $\hat{s}\ell^2\le 1$}.
\label{eq:fPE}
\end{equation}
Here $\Delta s$ is the vacuum expectation value difference and
$\ell =\ell_0/\gamma$ is the wall thickness, Lorentz-contracted from its value at rest $\ell_0$.
A similar result arises assuming instead a perfectly inelastic collision~\citeall.
Both lead to a function $f \sim 1/\hat{s}^2$  that extends up to large $\hat{s}=\gamma^2/\ell_0^2$.

\smallskip

We now argue that both factors in eq.\eq{frontaloff} are separately problematic: $\Im \Pi$ and the assumed $f_{\rm PE}$.

\subsection{The problem in the off-shell $\Pi$}
One problem with this approximation is that 
the full effective action $\Gamma[s]$ is physical when evaluated over a solution of the equations of motion $\delta\Gamma/\delta s = 0$.
Instead, when evaluated over a generic configuration, $\Gamma[s]$ depends on the field parameterisation by terms
proportional to the equations of motion.
The action approximated as its quadratic part and evaluated over an approximated field configuration is not physical.

This unphysical dependence manifests, in gauge theories, as gauge dependence, 
because different gauges describe the same physics in terms of different coordinates in field space, 
so that the gauge dependence of the effective action can be compensated by a  field redefinition~\cite{Nielsen:1975fs,Fukuda:1975di,Aitchison:1983ns,hep-ph/0501259}.

Gauge-dependence is a symptom of a deeper problem: the off-shell $\Pi(p^2)$ is not physical because off-shell amplitudes are not physical.
Under field redefinitions they change by terms proportional to the equation of motion $\delta \Gamma/\delta s$.
See~\cite{2312.06748} for a recent formal summary, and~\cite{2507.08803} for a concrete example,
that consists in performing the field reparametrizations $s\to s + g s^2$ in the free quadratic action for $s$ where $\Pi_0(p^2) = p^2 - m^2$.
On-shell amplitudes recognise that $s$ is a free scalar and vanish.
The unphysical off-shell propagator acquires the one loop correction
\beq \label{eq:prop1s}
\Pi(p) = \Pi_0(p) -2 g^2 \Pi_0(p) [6 A +\Pi_0(p) B(p) ]+\cdots\eeq
 where $A$ and $B$ are the usual Passarino-Veltman functions. 
Off-shell, $B$ provides unphysical imaginary parts.
The field redefinition introduced off-shell couplings.

\medskip

Since the production rate is inferred directly from the imaginary part of the off-shell propagator, 
its arbitrariness translates into an arbitrary final result, rather than being a harmless intermediate step.

This conclusion is consistent with the Nielsen identity. 
Schematically, for a gauge-fixing parameter $\xi$,
it says that the
value of the \emph{complete} effective action is gauge independent when evaluated on a stationary configuration 
of that same action, $\delta\Gamma/\delta s=0$:
\beq\label{eq:Nielsen}
 \frac{\partial \Gamma[s;\xi]}{\partial\xi}
 +\int d^4x\,K[s;\xi](x)\frac{\delta\Gamma[s;\xi]}{\delta s(x)}=0 .
\eeq
This is the mechanism that makes a consistently computed bounce action gauge independent.
It does not imply gauge independence if a background solving nonlinear classical equations is combined only with the absorptive quadratic kernel of a separately truncated quantum effective action.
Under a gauge variation, the omitted higher background insertions and the corresponding correction to the background are precisely among the terms required by eq.\eq{Nielsen}.
For a colliding wall their neglect is not controlled: 
the wall profile and its post-collision evolution exist because of order-one nonlinear self-dynamics.
Accordingly, the possible gauge dependence of the approximate background 
cannot in general cancel that of an isolated off-shell self-energy.

A nonlinear reparametrization of a free scalar provides a particularly sharp check.
It generates an imaginary part in an off-shell two-point function while all on-shell amplitudes remain zero.
Consequently, the cancellation cannot be automatic within eq.\eq{ImGamma}; the additional terms in the full effective action are indispensable.

\medskip

Eq.\eq{ImGamma} is presented in 
textbooks (e.g.~\cite{IZ}) to approximate particle production from electromagnetic fields $s\to A_\mu$
and in the literature to approximate particle production from 
gravitational fields $s\to h_{\mu\nu}$~\cite{Cespedes:1989kh,Campos:1991ff,2502.12249}.
Photons and gravitons are massless, so a non-vanishing particle production rate needs off-shell modes with `power spectrum' $|s_0(p)|^2$.
To clarify the issue, we now show that eq.\eq{ImGamma} gives the correct answer in a theory different from the problem at hand:
a free field $s$ perturbed by a classical current $J$, described by the action $S_2[s] =S_{0}[s] + \int d^4x\,  s J$
where $S_0 = \frac12 s \Pi_0 s $ is a free action, quadratic in $s$.
The non-trivial field configuration $s_0 $ is then sourced as $s_0 = -D J$ where $D = 1/\Pi $ is the dressed propagator.
In the photon case $J$ is the usual electro-magnetic current $J_\mu$;
in the graviton case $J$ is the energy-momentum tensor $T_{\mu\nu}$;
in the case of our scalar, its self-interaction needs to be mimicked by an effective source 
$J = (\partial^2 + m^2) s_0 =  - V'(s_0)$ able to produce the same profile as vacuum decay.
In all cases $J(p) =- \Pi(p) s_0(p)$.
The action $S_2[s]$ is quadratic, so its corresponding effective action is computed exactly in terms of
$W[J]=-\frac12 \int  J(-p) D(p) J(p) d^4p/(2\pi)^4$,
obtaining the particle production rate from its imaginary part:
\beq \label{eq:W}
N = 2\Im W[J] =-  \int \frac{d^4 p }{(2\pi)^4}\, |J(p)|^2 \Im D(p) = \int \frac{d^4 p}{(2\pi)^4}\, |s_0(p)|^2 \Im\Pi(p) = 2\Im \Gamma[s_0] \eeq
having used  $\Im D  =- (\Im\Pi)/|\Pi|^2$.
This discussion clarifies that eq.\eq{ImGamma} is applicable when the self-dynamics of $s$ is negligible.
Instead, the self-dynamics  of vacuum decay bubbles is non-perturbative in the scalar self-quartic $\lambda$.
So adding the imaginary part of  cubic and quartic terms of the effective action would not be enough to fix the formalism.
Non-perturbativity can be seen in multiple ways. 
First, the classical shift in the field at leading order in $\lambda$ is $\delta s/s \sim \lambda s^2/p^2$, which becomes of order unity 
for wave number $p \lesssim \lambda s^2 \sim m_s$.
Second, the classical equation can be rewritten in terms of rescaled variables 
$\tilde{s}=s/w$ and $\tilde{x}_\mu =m_s x_\mu$ such that it no longer depends on $\lambda$ and on $w$.
Third, from a quantum point of view, the expansion parameter is $n \lambda \sim 1$
where $n$ is the occupation number.
The situation simplifies in the ultra-relativistic limit $\gamma\gg1$, as discussed in section~\ref{sec:good}.



\medskip

To get the particle production  rate from $2\Im\Gamma$ one needs to compute the full effective action $\Gamma$ evaluated over a solution of the equations of motion. 

This computation is difficult but feasible when considering {\em one} vacuum decay bubble.
For example, vacuum decay from a possibly negative Higgs quartic was computed in~\cite{hep-ph/0104016}.
However the rate of particle production from one expanding bubble appears negligible~\cite{2403.03252}
compared to the rate from bubble collisions.
The classical field configuration $s_0(x)$ describing the collision is too complicated,
and it only solves approximated classical equations.

\smallskip

Assuming that the heavy particles are produced via a small coupling,  $\Im\Gamma$ can be expanded  in such coupling.
At leading orders this gives a reparametrization-invariant generalization of eq.\eq{ImGamma},
that contains the $s$ propagator evaluated over the collision background
\beq \label{eq:s0quantum} \bAk{s_0}{s(x) s(y)}{s_0} \approx s_0(x) s_0(y)+\cdots \eeq
rather than its classical limit as in eq.\eq{ImGamma}.
It is feasible to compute the extra quantum contribution when the background field is time-dependent only,
by identifying the physical fluctuations around  $s_0(t)$.
However, eq.\eq{s0quantum} remains a formal statement when it needs
to be computed on a too complicated field describing a wall collision.
We will thereby move to a different approach.

\begin{figure}[t]
$$\includegraphics[width=\textwidth]{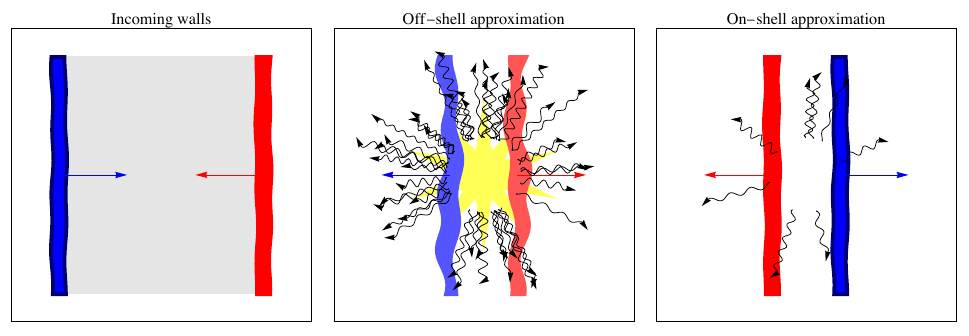}$$
\begin{center}
\caption{\it\label{fig:CollidingWallsCartoon} 
The left panel shows two incoming bubble walls, in their center-of-mass frame.
The other panels show a schematic comparison of the two pictures.
According to~\cite{1104.4793,1211.5615,1608.00583,2308.13070,2308.16224,2403.03252,2407.16747,2506.12123,2412.17912,2512.13815,2601.02458,2605.03758} the walls bounce back, sending all their quanta
far off-shell, allowing abundant heavy particle production (middle panel). 
In our computation the ultra-relativistic walls undergo nearly free passage to leading order, 
while hard production arises only from rare microscopic partonic scatterings (right panel).
}
\end{center}
\end{figure}

\subsection{The problem in the collision factor $f_{\rm PE}$}\label{sec:problemgamma}
The approximation in eq.\eq{fPE}, according to which the two colliding walls reflect off each other
in a complicated way, approximated as either elastically or inelastically, 
is qualitatively incorrect in the ultra-relativistic regime $\gamma\gg1$. 
In this limit the physics is simple: 
in the center-of-mass frame the two Lorentz-contracted walls overlap for a time $\sim\ell_0/\gamma$,
parametrically shorter than the time scale on which the potential force can significantly deflect the fields.
As a result, the leading collision profile is given by nearly-free passage, i.e.~by the approximately linear superposition of the two incoming walls, 
up to corrections suppressed by the Lorentz factor,
$\delta s/s\sim 1/\gamma^2$~\cite{Hawking:1982ga,1005.3493,1906.02588}.
So the walls initially largely pass through each other rather than bounce, 
producing a transient intermediate region in which the field is displaced away from the minima.
See fig.\fig{CollidingWallsCartoon} for an illustration.

The potential force later drives rollback and re-collisions on a timescale of order $1/m_s$.
Such nonlinear aftermath can be important for reheating and can produce scalar quanta at invariant masses of order $m_s$.
The field configuration contains large Fourier momenta $k,\omega \sim\gamma m_s$ only in the free walls
relative to the rest frame;
no longer in the combination $ \hat{s} \equiv \omega^2-\mathbf{k}^2$
that can lead to production of heavy particles up to $M \lesssim \hat{s}/4 \gg m_s$.
This distinction can be seen from the explicit solution for a scalar with potential $V(s)$ 
\beq \delta  s =-V'(2 s_{\rm true} - s_{\rm false})(t^2-x^2)/4 + \cdots \eeq
at leading order in $t,x$, inside the future light cone of the collision.
Thus the departure from free passage starts continuously and is quadratically suppressed.
At the first-overlap time $t\sim\ell_0/\gamma$ it is of relative order $1/\gamma^2$.
Its Fourier transform gives a soft $f(\hat s) \propto V^{\prime 2}/\hat{s}^4$, parametrically suppressed at $\hat s \gg m_s^2$
compared to eq.\eq{fPE}.

\medskip

The hard effect claimed in~\citeall{} may arise from the following issues.
First, a perfectly elastic collision looks like no collision, if the identity of the two walls is confused.
Indeed, choosing the field origin such that the false and true vacua lie at $0$ and $\Delta s$, respectively, 
the Free Passage and Perfectly Elastic configurations for a sharp wall are
\beq
s_{\rm FP}=\Delta s[\Theta(t-x)+\Theta(t+x)],
\qquad
s_{\rm PE}=\Delta s\,  \Theta(|x|-|t|).
\eeq
Although their unlabelled wall trajectories coincide, their field configurations do not: 
the field between the outgoing walls is $2\Delta s$ after free passage and $0$ after reflection. 
Confusing the two introduces the extra term
$
s_{\rm PE}-s_{\rm FP}=-2\Delta s\Theta(t-|x|),
$
whose Fourier transform is proportional to $\Delta s/p^2$, generating the artificial hard collision tail.
This confusion might happen by assuming a field ansatz that depends on $|t|$ rather than $t$ where $t=0$ is the collision time
(see e.g.\ eq.~(2.3) of~\cite{1211.5615})
and/or because the collision instant is not resolved by numerical simulations in thin-wall approximation.
Second, even the Fourier transform of a no-collision configuration acquires off-shell hard quanta, 
if the slow time-scale of the rollback force is confused with the fast time-scale of the collisions,
when e.g.\ the difference between an elastic and an inelastic collision is assumed to reach short times.
Third, the wall collision can be computed analytically for the sine-Gordon model with $\sin(s/f)$ potential~\cite{Bowtell:1977yh}
and shows the expected ultra-relativistic suppression.
This was noticed in~\cite{Watkins:1991zt} but dismissed as a peculiarity of the integrable model.

\smallskip

The issue of correctly understanding the collision has mild consequences also for gravitational waves produced during the later phase, better approximated by the `bulk flow' method  rather than by the `envelope' method,
as already discussed in the literature~\cite{1707.03111,1712.06869}.

\section{On-shell approximation}\label{sec:good}
Hard quanta and/or heavy particles are instead dominantly produced at the first impact of the two Lorentz-contracted plane waves.
We approximate the first impact  by treating each wall as a coherent state of on-shell quanta of the scalar field $s$, 
and by computing ordinary on-shell scattering processes among these quanta.

To see that this different approach is necessary, let us assume that
 $s$ couples as $ \lambda_{s\phi}s^2\phi^2/2$ to some extra scalar $\phi$.
Inserting the leading free-passage background $s=s_L+s_R$ 
into the interaction generates the cross term $s_Ls_R$.
Although $s_L$ and $s_R$ separately have space-like Fourier support, their product has hard time-like support.
Expanding the two incoming walls into modes and applying the standard LSZ formalism 
gives the on-shell scattering $ss\to \phi\phi$ considered below.
This contribution is absent from eq.~\eq{ImGamma}, as its rate is truncated to quadratic order in $s$. 
The truncation omits the physical process that is dominant in the hard limit.
Applying the off-shell approach with the correct $f_{\rm PE}$ would not capture this effect.
Our new approach is necessary.

This approach is analogous to the  Weizsäcker-Williams approximation of an electromagnetic field in terms of quasi-real equivalent photons,
and to the related parton model of high-energy hadron collisions.
Hard particle production from proton-proton $pp$ collisions is not approximated by
viewing the protons as two balls that collide in a perfectly elastic or inelastic way
sending all quarks and gluons far off-shell.
Rather, each proton is approximated as a collection of partons and one computes collisions of partonic constituents.
This approximation works because the protons are highly Lorentz contracted, so that 
their different parts don't have enough time to interact.
Similarly, the first wall collision can be approximated as a partonic-like set of collisions.

Of course, a classical wall is not an incoherent gas of particles.
The on-shell partonic treatment neglects coherence effects between collective partonic sub-collisions.
As usual, such coherence effects get sub-leading for the hard processes of interest, 
with  large enough invariant energy $\sqrt{\hat s}\gg m_s$ exchanged in a given microscopic collision.
This approximation is relevant for the production of particles much heavier than $m_s$, that dominantly arise at the first splash.

\smallskip

We consider one planar wall moving along the $x$ axis.
We denote it as $s_0(t,x)$, despite that in the previous section the same symbol denoted two colliding walls.
The wall profile can be conveniently computed in uniformly accelerated Rindler coordinates $ds^2 = -(a x')^2 d\tau^2 +dx'^2$ where
it is the static  solution
\beq s_0(x,t)=s_0(x') = s_0 (\sqrt{x^2-t^2})\eeq 
 to the classical equation 
\beq \frac{d}{dx'}\bigg[\frac12\bigg(\frac{ds_0}{dx'}\bigg)^2-V\bigg] = - \frac{1 }{x'} \bigg(\frac{ds_0}{dx'}\bigg)^2.\eeq
However, the wall proper acceleration $a$ contributes to particle production much less than wall collisions~\cite{2403.03252}.
For simplicity we neglect $a$ and focus on the two wall configurations just before their collision.
Then, neglecting the acceleration and thereby the potential energy difference between the two vacua, the static solution $s_0(x')$ more
simply arises in the rest frame $x'$ of the wall, where it solves the classical equation 
$ds_0/dx' = \pm \sqrt{2V}$.
The wall profile can be Fourier expanded as $s_0(x') = \int (dk/2\pi) e^{ikx'} s_0(k)$ with $s_0(-k)=s_0^*(k)$.
The wall energy density is 
\beq \sigma = \int dx'[K+V]=  \int dx' \left (\frac{ds_0}{dx'}\right)^2=\int_0^{\infty} \frac{dk}{\pi} \, k^2 |s_0(k)|^2.\eeq
Neglecting the wall proper acceleration $a= \Delta V/\sigma$,
a right-moving planar wall is then simply approximated by boosting the quasi-static wall,
\beq s_R(x,t) = s_0 [\gamma(x-v t)].\eeq
In the ultra-relativistic limit this can be represented as a coherent state of quasi-real quanta 
of the scalar field $s$
with momentum $p_\mu = (E_R, p_R,0,0)$ and positive $p_R $:
\beq
s_R(x)=
\int _0^\infty\frac{dp_R}{2\pi \sqrt{2E_R}}
\left[
\alpha(p_R )e^{-ip\cdot x}+\alpha^*(p_R) e^{ip\cdot x}
\right].
\eeq
The full scalar field is still canonically quantized over all momenta.
The space-like virtuality at rest gets suppressed by $1/\gamma^2$  in the ultra-relativistic limit 
$E_p=\sqrt{p_R^2+m_s^2} \simeq p_R$.
The field coefficients $\alpha$ can be expressed in terms of the Fourier transform of wall at rest as
\beq \alpha(p_R)  = \frac{\sqrt{2 E_R}}{\gamma} s_0\left(\frac{p_R}{\gamma}\right).\eeq
The occupation number of $s$-quanta and the energy per unit transverse area $A$  are
\beq
\frac{N_R}{A}=\int_0^\infty \frac{dp_R}{2\pi }|\alpha(p_R)|^2 = \int _0^\infty dp_R F_R(p_R),
\qquad
\frac{E_{R\rm wall} }{A} =
\int_0^\infty \frac{dp_R}{2\pi}E_p|\alpha(p_R)|^2 .
\eeq
We here introduced the wall parton density per unit transverse area 
\beq
F_R(p_R) \equiv
\frac{1}{A}
\frac{dN_R}{dp_R}= \frac{p_R}{\pi \gamma^2} \left| s_0\left(\frac{p_R}{\gamma}\right) \right|^2 .
\eeq
For a planar wall with tension $\sigma$ boosted to Lorentz factor $\gamma \gg 1$
\beq
\frac{E_{R\rm wall}}{A}\simeq \frac{P_{R\rm wall}}{A} \simeq \gamma\sigma.
\eeq
We next compute scatterings among two quanta in their center of mass rest frame,
where the left-moving wall has the same Lorentz factor $\gamma$.
The invariant mass of an ultra-relativistic pair of colliding wall partons is 
$
\hat s=(p_L+p_R)^2_\mu
\simeq 4 p_R p_L
$.
The differential luminosity per unit transverse area is then
\beq
\frac{d{\cal L}_{ss}}{d\hat s} =
\int dp_L dp_R\, F_L(p_L) F_R(p_R) \delta(\hat s- 4 p_L p_R).
\eeq
In terms of rest-frame momenta $k_{L,R}$ and of the rest-frame Fourier transform $s_0(k)$ it is
\beq 
\frac{d{\cal L}_{ss}}{d\hat s}=\int \frac{dk_L}{\pi}\frac{dk_R}{\pi} k_L k_R |s_0(k_L)s_0(k_R)|^2 \delta (\hat s - 4 \gamma^2 k_L k_R)
=\frac{\hat{s}}{16 \pi^2 \gamma^4}\int dr \bigg|s_0(\frac{\sqrt{\hat s}}{2\gamma}e^r) s_0(\frac{\sqrt{\hat s}}{2\gamma}e^{-r}) \bigg|^2\eeq
where $r=\ln \sqrt{k_L/k_R}$.
The  number of  hard collisions producing particles $f$ is then computed in terms of partonic cross sections $\hat\sigma$ as
\beq
\frac{N_f}{A}
=
\int d\hat s   \frac{d{\cal L}_{ss}}{d\hat s}
\hat\sigma_{ss\to f}(\hat s)=
\int \frac{dk_L}{\pi}\frac{dk_R}{\pi} k_L k_R |s_0(k_L)s_0(k_R)|^2 \hat\sigma_{ss\to f}(4 \gamma^2 k_L k_R).
\eeq
The factor ${\cal L}_{ss}$ has mass dimension 4, because it's a luminosity per unit area.
The formula generalizes in the usual way if multiple particle species $a,b$ are present in the wall field,
or if particles from the cosmic bath accumulate around the walls~\cite{2403.05615}.
We here assume one real scalar $s$, such that the  collision ultimately leaves the true vacuum on both sides.
The on-shell partonic cross sections are gauge-invariant and physical.
In the theory considered here the wall is described by a canonically normalized physical real scalar, so its classical profile and the resulting luminosity contain no gauge-fixing parameter.
For a wall generated by a gauge-Higgs system beyond tree level, the split into a constituent distribution and a short-distance coefficient can depend on the interpolating field or factorization convention.
The physical convolution must then be matched to canonically normalized asymptotic states, or formulated using gauge-invariant operators, with the background computed consistently at the same order.
This standard qualification does not rescue an isolated gauge-dependent off-shell subprocess: in the on-shell calculation the required cancellations occur among the complete set of contact and exchange diagrams.

Fig.\fig{FeynhhZZ} illustrates the Feynman diagrams that contribute to vector production.
As discussed in section~\ref{sec:comparison},
the $s$-channel diagram provides an off-shell $ss \to s^*$ that resembles the off-shell approach.
It is gauge-dependent and completed by other diagrams into a physical gauge-invariant cross section, often dominated by the
other diagrams not present in the off-shell approach.


\subsection{Computing the wall profile}
\label{III}
To proceed and make the previous discussion concrete we need to specify a theory.
We consider a scalar with a potential $V(s)$ with two minima arising from
a self-quartic $\lambda$. In the thin-wall limit it can be written as
\beq \label{eq:Vs}
V(s) = \frac{\lambda}{4} (s^2 - w^2)^2 + \Delta V \frac{s}{2w}\eeq
for small $\Delta V$, such that the scalar mass is $m_s=\sqrt{2\lambda} w$.
 The bubble energy grows because it accumulates the vacuum energy in its past volume.
 Large $\gamma \gg 1$ can be reached at collision, as will be summarised in section~\ref{sec:gamma}.
Neglecting the wall acceleration, the field equation  $d^2 s_0/dx'^2 = V'$
has the first integral $ds_0/dx' = \pm \sqrt{2V}$, solved by
\beq x'(s) =\int^s \frac{ds'}{\sqrt{2V(s')}}.\eeq 
Near the two minima one can approximate $V\simeq m_s^2 (s-\med{s})^2/2$ 
with $m_s = w\sqrt{2\lambda}$, obtaining 
$s -\med{s}\propto e^{\pm m_s x'}$.
Neglecting $\Delta V$ in eq.\eq{Vs} gives
\beq \label{eq:s0tanh}
s_0(x')=w \tanh\bigg [\frac{m_s}{2} x'\bigg]\eeq 
for a wall centred around $x'=0$.
The wall thickness is $\ell_0 = 2/m_s$ in the convention of~\cite{1211.5615}, and the wall tension is $\sigma=\frac23 m_s w^2$.
Taking $\Delta V$ into account gives qualitatively similar profiles with modified values of the wall thickness. 
The dimension-less Fourier transform is
\beq s_0(k) \equiv \int dx\,e^{-ik x'} s_0(x') = - 2\pi i \frac{w}{m_s}\operatorname{csch} \frac{\pi k}{m_s} \simeq 
\left\{ \begin{array}{ll}
- 2i w/k & k\ll m_s,\cr
-4\pi i (w /m_s) e^{-\pi k/m_s} & k\gg m_s .
\end{array}\right.
\eeq
The Fourier transform of the vacuum jump gives the $1/k$ tail, starting from $k \sim m_s$.
This is a general feature: the Fourier transform of any $s_0(x')$ that solves the equation of motion obeys
\beq s_0(k) = \frac{1}{ik} \int ds~e^{-i k x'(s)}.\eeq
This IR behaviour is well known in QFT,
e.g.\ the partonic distribution of any massive charged particle contains a similar log-divergent tail of soft vectors,
cut by the particle masses.
This tail carries little energy, and does not contribute to the highest-energy events.


\begin{figure}[t]
$$\includegraphics[width=0.6\textwidth]{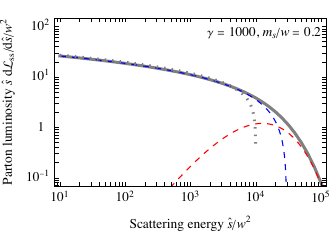}$$
\begin{center}
\caption{\it\label{fig:WallPartonLumi} Example of wall partonic luminosity $\hat s\, d{\cal L}_{ss}/d\hat s$
(gray) compared to its low-energy (blue) and high-energy (red) asymptotics.
The dotted curve is the parton luminosity equivalent to the problematic off-shell approach, 
arbitrarily assuming $\Im \Pi (\hat s)=2 \hat \sigma\hat s w^2$, see eq.\eq{OffLumi}.}
\end{center}
\end{figure}


\subsection{Computing the wall parton luminosity}
The above discussion shows that the dimension-less Fourier transform of a typical wall profile is $s_0(k) \sim w/k$ up to $k \lesssim m_s$.
The occupation number is $n_k \sim |s_0(k)|^2 \sim 1/\lambda$ at $k \sim m_s$. 
So the number of quanta in the wall receives a logarithmic contribution from soft modes,
$N_s/A \approx  w^2 \ln m_s/k_{\rm min}$.
On the other hand, the wall energy is dominated by hard quanta.
The parton luminosity is roughly constant up to $\sqrt{\hat{s}}\sim \gamma m_s$:
\beq \hat s   \frac{d{\cal L}_{ss}}{d\hat s} \sim w^4 \ln \frac{k_{\rm max}}{k_{\rm min}}.\eeq
Configurations with asymmetric $k_L/k_R$ contribute to the log tail with $k_{\rm max}\sim m_s$ and $k_{\rm min} \sim \hat{s}/m_s\gamma^2$.
A more precise approximation is obtained considering the wall profile of eq.\eq{s0tanh}, with thickness $\ell_0 = 2/m_s$
\beq \label{eq:partonlumi}
\hat s   \frac{d{\cal L}_{ss}}{d\hat s}  =\frac{\pi^2 w^4}{\gamma^4 m_s^4}\hat{s}^2 J(\frac{\pi\sqrt{\hat s}}{2\gamma m_s})
\simeq  \frac{32 w^4}{\pi^2} \left\{\begin{array}{ll}
\displaystyle  \ln \frac{\gamma e m_s}{\pi\sqrt{\hat s}}  & \sqrt{\hat{s}}\ll \gamma m_s\cr
\displaystyle \frac{\pi^4\hat{s}^{7/4}}{2 (m_s \gamma)^{7/2}} e^{-2\pi \sqrt{\hat s}/m_s\gamma}
& \sqrt{\hat{s}}\gg \gamma m_s\
\end{array}\right.
\eeq
where 
$J(z) =\int_{-\infty}^{+\infty}  dr \operatorname{csch}^2(z e^{-r})\operatorname{csch}^2(z e^{+r}) $ and we wrote its asymptotic for small and large argument.
 Fig.\fig{WallPartonLumi} shows the numerical result for the parton luminosity,
comparing it to the small-$\hat s$ and large-$\hat s$ limit.
The log term at small-$\hat{s}$ comes from asymmetric configurations, and draws at least one leg from the coherent, highly occupied tail.
So saturation-type corrections are expected to be significant and could be computed similarly to
the dilute-dense regime of QCD at small $x$, see e.g.~\cite{hep-ph/0303204}.
In the special case of the tanh wall, the potential for fluctuations is the reflection-less P\"oschl-Teller well,
so that saturation might induce  phase shifts only.
We proceed by ignoring such effects.
The logarithmic enhancement at low $\hat s$ is therefore an estimate of the soft tail rather than a precision prediction. 
Our main phenomenological conclusions concern heavy-particle production in the hard regime, 
where the crucial difference from the off-shell approach is the absence of an unsuppressed hard collective source 
rather than the precise normalization of the soft tail.

\subsection{Comparing the on-shell and the off-shell results}\label{sec:comparison}
Assuming the efficiency factor $f_{\rm PE}$ of eq.\eq{fPE}~\citeall, 
the off-shell approach gives the production rate 
\beq \frac{N}{A} \stackrel{\hat{s}\ell^2\le 1}\simeq \frac{32w^2}{\pi^2}\int \frac{d\hat{s}}{\hat{s}} \frac{\Im \Pi}{\hat{s}} \ln \frac{\gamma^2 m_s^2}{\hat s}.\eeq
This has a logarithmic form, similar to the low-energy tail of eq.\eq{partonlumi}.
This helps to compare the off-shell to the on-shell rates with  eq.\eq{partonlumi}.
They are equal provided that 
\beq\label{eq:OffLumi} \frac{\Im \Pi(\hat{s})}{2\hat{s}\hat\sigma(\hat{s}) w^2} \sim 1.\eeq
In general, such factor is model dependent, as $\Im \Pi$ and $\hat\sigma$ are different objects.
In section~\ref{sec:cross} we will compute and compare the main cases.
We here estimate a typical dimension-less coupling $g$, that
gives a typical cross section $\hat\sigma \sim g^4/\hat{s}$ and a typical $\Im\Pi \sim \hat{s} g^2$.
Therefore
\beq \frac{\Im \Pi}{\hat{s}\hat\sigma w^2} \sim \frac{\hat{s}}{g^2 w^2}.\eeq
This means similar rates in the soft regime $\hat{s}\sim g^2 w^2$, 
while {\em the off-shell approach gives parametrically larger rates than the on-shell approach
in the hard regime} $\hat {s}\gg g^2 w^2$.
Roughly speaking, the on-shell approach gives the usual hard-scattering behaviour:
high energies imply small lengths and thereby small cross sections.
Thus only a small fraction of the wall constituents participate in hard collisions, just as only a small fraction of partons in a high-energy proton-proton 
collision undergo a hard scattering.
The off-shell approach, instead, assumes in eq.\eq{fPE} that collective field dynamics is
an unsuppressed source of hard time-like modes.
As we argued in section~\ref{sec:problemgamma}, classical solutions behave differently: ultra-relativistic walls freely cross each other, up to $\gamma^2$-suppressed
corrections.

\smallskip

To clarify the situation, we focus on one contribution present in the on-shell formalism that
can be more directly compared with the off-shell formalism.
In the off-shell picture, the wall collision itself is assumed to generate a hard time-like virtual scalar. 
In the on-shell picture, a virtual scalar can arise as the mediator of an ordinary microscopic scattering
$ss \to s^* \to f$.
This happens in the first Feynman diagram in fig.\fig{Feynssff} or the second diagram in fig.s\fig{Feynssphiphi} and\fig{FeynhhZZ}.
In this case, both approaches may be viewed as providing an effective distribution of virtual $s^*(k)$ modes.
The off-shell approach assumes with eq.\eq{fPE} that the collective rebounce of the colliding walls drives all $s^*$ quanta off-shell.
By contrast, the on-shell approach derives it from scatterings of quasi-real wall quanta,
giving a rate controlled by physical on-shell amplitudes proportional to the scalar couplings.
This special scattering amplitude factorises as
$$
\mathscr{A}_{ss\to f}
=
\mathscr{A}_{ss\to s^\ast} D(\hat s)
\mathscr{A}_{s^\ast\to f}
$$
where $D(\hat s)={i}/(\hat s-m_s^2+i m_s\Gamma_s)$ is the $s$ propagator and 
the scalar cubic is $\mathscr{A}_{ss\to s^\ast} = 3 m_s^2/w$ for the quartic potential of eq.\eq{Vs}. 
It is super-renormalizable, so
the resulting partonic cross section is suppressed at $\hat s \gg m_s^2$, where the on-shell approach is applicable: 
\beq
\hat\sigma(ss\to f)
\simeq
 \frac{\mathscr{A}_{ss\to s^\ast} ^2}{\hat{s}^{5/2} }\Gamma(s^* \to f),\qquad
\Im \Pi = \sqrt{\hat s} \Gamma(s^* \to f),
\qquad
\frac{\Im \Pi}{2\hat{s}\hat\sigma w^2} =\frac{\hat{s}^2}{18 m_s^4}.\eeq
The two approaches roughly agree in the soft regime $\hat{s}\sim m_s^2$, but strongly differ in the hard regime $\hat s\gg m_s^2$.
A different scattering process, $ss \to s s^*$ followed by $s^*$ decay, avoids the super-renormalizable suppression.
But its hard rate remains parametrically lower than in the off-shell approach.

This suppression matches the $1/\gamma^2$ suppression of ultra-relativistic classical wall collisions
discussed in section~\ref{sec:problemgamma}.
After correcting the off-shell approach removing the hard $1/\hat{s}^2$ tail from eq.\eq{fPE}~\citeall, 
the off-shell description can capture soft coherent modes for which the incoherent partonic treatment is inappropriate.
The on-shell approach is appropriate for hard modes, above a matching scale of order $m_s$.
This is similar to proton-proton collisions at $\sqrt{\hat s}\gg \Lambda_{\rm QCD}$:
the parton approach is appropriate for rare but interesting high-energy scatterings,
while the collective dynamics controls soft modes, around the QCD scale.

\section{Partonic cross sections}\label{sec:cross}
We here compute the partonic cross sections assuming the simplest interactions of the wall scalar $s$ to itself, to other scalars,
to fermions, to vectors.
In each case we emphasize the threshold behaviour, the hard-energy asymptotic, and the comparison with the corresponding off-shell estimate.

\subsection{Scalar self-production}
The self-scattering amplitude  is
\beq \mathscr{A}(ss\to ss) = 6i\lambda \left(1 + \frac{3 m_s^2}{\hat s-m_s^2} +\frac{3 m_s^2}{ \hat t-m_s^2} +\frac{3 m_s^2}{\hat u-m_s^2} \right).\eeq
The cross section is $\hat \sigma \simeq 9 \lambda^2/8\pi \hat s$ 
in the hard regime $\hat{s}\gg m_s^2$, where the $s$-channel diagram is negligible.
This means that wall collisions produce  $N_s/A \sim m_s^4/\hat{s}$ quanta with energy $\hat{s}$.


\medskip

This can be compared to the off-shell approach, where  $\Im\Pi \sim \lambda^2 w^2/4\pi  + \lambda^2\hat s/(4\pi)^3$~\cite{2403.03252}.
The first term is suppressed by the super-renormalizable coupling, 
and gives a production rate comparable to that obtained in the on-shell formalism,
since $\Im\Pi /\hat{s}\hat \sigma w^2 \sim 1$.
The agreement is accidental.
The second term is not subject to the super-renormalizable suppression. 
It arises from the three-body off-shell decay $s^*\to sss$~\cite{2403.03252}, and leads to an enhancement at large $\hat s$. 
As discussed in section~\ref{sec:comparison} no corresponding enhancement occurs in the on-shell formalism.
The two results are qualitatively different.
The off-shell approach incorrectly assumes that some collective effect
makes the two walls unable to cross each other sending all their quanta maximally off-shell.

 \begin{figure}[t]
$$\includegraphics[width=\textwidth]{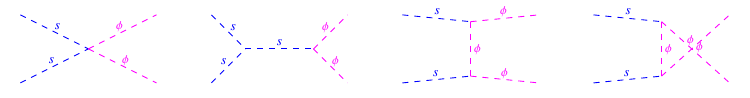}$$
\begin{center}
\caption{\em\label{fig:Feynssphiphi}\em Feynman diagrams for scalar $\phi$ pair production from collisions of two scalars $s$.}
\end{center}
\end{figure}
 
\subsection{Scalar production}
We next consider the production of an extra stable scalar $\phi$ quadratically coupled to $s$ in the potential 
 \beq V(s,\phi)= V(s) + \frac{\bar{m}_\phi^2}{2} \phi^2 + \frac{\lambda_{s\phi}}{4} s^2 \phi^2 + \cdots,\eeq
 so that $m_\phi^2 = \bar{m}_\phi^2 + \lambda_{s\phi} w^2/2$ is the scalar mass in the broken phase.
The Feynman diagrams in fig.\fig{Feynssphiphi} give the scattering amplitude
\beq \mathscr{A}(ss\to \phi\phi) = i\lambda_{s\phi} \left(1 + \frac{3 m_s^2}{\hat s-m_s^2} +\frac{\lambda_{s\phi} w^2}{ \hat t-m_\phi^2} +\frac{\lambda_{s\phi} w^2}{\hat u-m_\phi^2} \right).\eeq
The second diagram is negligible in the hard limit $\hat s\gg m_s^2$, where the cross section is 
\beq\begin{split}
\hat\sigma(ss\to \phi\phi) &= \frac{\int dt\, |\mathscr{A}|^2}{32\pi \hat{s}^2}
=\frac{\lambda_{s\phi}^2 }{32\pi \hat s}R-
\frac{\lambda_{s\phi}^3 w^2 }{4\pi\hat{s}^2}{\rm arctanh} \, R+
\frac{\lambda_{s\phi}^4 w^4}{4\pi\hat s^3}
 \bigg({\rm arctanh}\, R+\frac{R}{1-R^2} \bigg)
\end{split}
\eeq
where $R = \sqrt{1-4m_\phi^2/\hat{s}}$ is the $\phi$ velocity.

 \begin{figure}[t]
$$\includegraphics[width=0.6\textwidth]{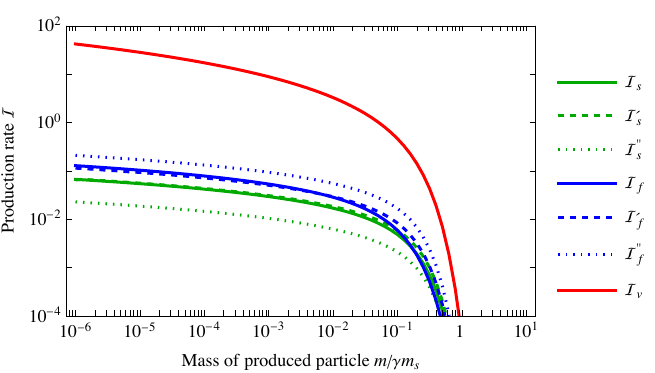}$$
\begin{center}
\caption{\it\label{fig:WallProductionRates} The functions ${\cal I}_s$, ${\cal I}_f$, ${\cal I}_V$ that control as in eq.\eq{Is},\eq{If},\eq{IV}
the production rates of scalars, fermions, vectors 
with mass (here generically denoted as $m$) from collisions of walls made of scalars with mass $m_s$ boosted to a Lorentz factor $\gamma \gg 1$.}
\end{center}
\end{figure}

In the limit $\hat{s}\gg m_\phi^2$ the purely quartic diagram dominates and 
$\hat\sigma \simeq {\lambda_{s\phi}^2}/{32\pi \hat{s}}$.
Near threshold, by contrast, the $t$- and $u$-channel exchange terms can be relevant when a significant fraction of $m_\phi^2$ arises from symmetry breaking, so the threshold normalization need not track the asymptotic.
The partonic integral gives
\beq \label{eq:Is}
\frac{N_{\phi\phi}}{A}= \frac{w^4}{m_\phi^2 } \bigg[   \lambda_{s\phi}^2 {\cal I}_s(\frac{m_\phi}{\gamma m_s})  -
 \lambda_{s\phi}^3\frac{w^2 }{m_\phi^2} {\cal I}'_s(\frac{m_\phi}{\gamma m_s}) +
  \lambda_{s\phi}^4\frac{w^4 }{m_\phi^4}  {\cal I}''_s(\frac{m_\phi}{\gamma m_s}) 
\bigg]\eeq
where the dimension-less ${\cal I}$ functions are plotted in fig.\fig{WallProductionRates} and given, for example, by 
\beq
{\cal I}_s(r)=\frac{\pi}{8} r^4\int_1^\infty dx\, x \, J(\pi r\sqrt{x}) \sqrt{1-\frac{1}{x}} ,\qquad \hat{s} = 4 m_\phi^2 x.
\eeq

\medskip

This can be compared to the off-shell approach, that gives~\cite{1211.5615,2403.03252}
\beq \label{eq:Piscalar}\Im \Pi (\hat s)=\lambda_{s\phi}^2 \left(\frac{ w^2}{2\pi}+\frac{\hat s}{1024\pi^3}\right) \sqrt{1 - \frac{4 m_\phi^2}{\hat{s}}}.\eeq
According to~\cite{2403.03252} the second term arises from a three-body $s^*\to s\phi\phi$ decay.
If the three-body term is negligible, the on-shell and off-shell results are accidentally comparable around threshold.

 
 \begin{figure}[t]
$$\includegraphics[width=0.85\textwidth]{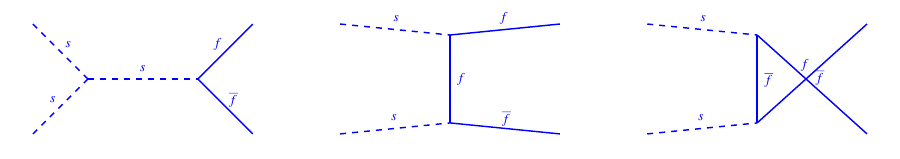}$$
\begin{center}
\caption{\em\label{fig:Feynssff}\em Feynman diagrams for fermion $f$ pair production from collisions of two scalars $s$.}
\end{center}
\end{figure}

\subsection{Fermion production}
We consider the production of a Dirac fermion $f$ coupled to the scalar $s$ by a Yukawa interaction $y s\bar{f} f/\sqrt{2}$.
This Yukawa contributes to the fermion mass as $m_f = \bar{m}_f +y w /\sqrt{2} $ where $\bar{m}_f$ is a possible mass term unrelated to symmetry breaking.
Fig.\fig{Feynssff} shows the Feynman diagrams that contribute to the partonic cross section.
The total cross section is
\beq \hat\sigma (ss\to f\! \bar f)= 
\bigg(\frac{y^4 }{8 \pi  \hat{s} } +\frac{3 \lambda y^3   w m_f }{\sqrt{2}\pi \hat s^2}\bigg) 
({\rm arctanh} \, R- R) +
\frac{9 \lambda^2 y^2 w^2 }{4\pi\hat{s}^2} R^3
 ,\qquad R=\sqrt{1-4 \frac{m_f^2}{\hat s}}.
\eeq
The cross section is halved if $f$ is a Majorana fermion, rather than a Dirac fermion.
At $\hat s \gg m_s^2$ it is dominated by the purely Yukawa contribution.
Near threshold at $R\ll 1$ one has $ {\rm arctanh} \, R- R \simeq R^3/3$ giving the expected phase space  suppression.
The first Feynman diagram involving off-shell $s^*$ exchange is suppressed at high energy
and is relevant near threshold if the quartic scalar is large, $\lambda \gtrsim y m_f/w$.
Integrating over the parton density gives the production rate
\beq \label{eq:If}\frac{N_{f\!\bar f}}{A}=\frac{w^4}{m_f^2} 
\bigg[ y^4  {\cal I}_f(\frac{m_f}{\gamma  m_s})+
y^3 \lambda \frac{w}{m_f}  {\cal I}'_f(\frac{m_f}{\gamma  m_s})+
 \lambda^2 \frac{y^2 w^2}{m_f^2}   {\cal I}''_f(\frac{m_f}{\gamma  m_s})\bigg]
\eeq
where the dimension-less functions ${\cal I}_f$ are plotted in fig.\fig{WallProductionRates}.
Furthermore, the 3-body process $ss\to s f\bar f$ would contribute as $\lambda^2 y^2$ without a suppression at $w \ll m_f$.

\medskip

The off-shell approach computes instead the $s^* \to ff$ rate obtaining
$ \Im \Pi (\hat s)= \hat{s} {y^2R^3}/{16\pi}$~\cite{2403.03252}.
According to eq.\eq{OffLumi} the ratio between the rates predicted by the two approaches is controlled by
${\Im \Pi}/{\hat{s}\hat\sigma w^2} \sim  \hat{s}/(yw)^2$.
The two results are comparable near threshold at $\hat{s}\sim m_f^2$, if the fermion mass arises entirely from the Yukawa coupling, $m_f \approx y w$ and $y \sim 1$.
However, even in this case, the off-shell approach predicts a growing rate of higher-energy fermion pairs.


\begin{figure}[t]
$$\includegraphics[width=\textwidth]{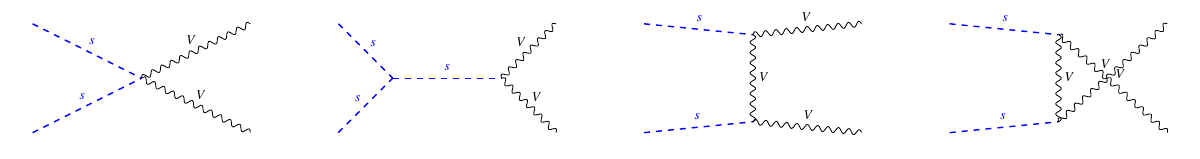}$$
\begin{center}
\caption{\em\label{fig:FeynhhZZ}\em Gauge-invariant set of Feynman diagrams for vector $V$ production from collisions of two scalars $s$.}
\end{center}
\end{figure}

\subsection{Vector production}
We next consider the production of a massive vector $V_\mu$.
Unlike in the previous cases, we focus on a theory where the vector mass $M_V$ arises {\em entirely} from 
its gauge coupling $g$ to a Higgs multiplet $S$ whose radial component is the physical $s$ component. 
Generalising this to theories where $M_V$ has extra sources requires additional scalars, which open extra scattering processes.
For simplicity, we assume a U(1) gauge group broken by a singlet $S$ with unit charge,  
 that acquires the same phase in both bubbles.
The action
\beq  \Lag = - \frac{1}{4}V_{\mu\nu}^2 +  |D_\mu S|^2 - V(S),\qquad
V(S)
=
\lambda\left(|S|^2 -\frac{w^2}{2}\right)^2 \eeq
gives, after symmetry breaking, in the unitary gauge:\footnote{Imposing the dark charge-conjugation symmetry
$V_\mu\to -V_\mu$,
$S\to S^*$
forbids $V_\mu$ kinetic mixing with hypercharge, thereby making the vector a stable DM candidate.
A more elegant Dark Matter theory is obtained assuming a gauge $\SU(2)$ broken by a scalar doublet~\cite{0811.0172,1306.2329},
to be studied in section~\ref{sec:2models}.}
\beq
S =\frac{w+s}{\sqrt{2}},
\qquad
M_V =gw,\qquad
m_s = \sqrt{2\lambda}w .\eeq
The vector couples to $s$ as
$ g^2 (s  w  + s^2  /2 )V_\mu^2$.
The first interaction is relevant in the off-shell approach.
Both interactions contribute to the physical $ss\to VV$ amplitude,
given  in the unitary gauge by the four Feynman diagrams in fig.\fig{FeynhhZZ}.
The second diagram contains the gauge-dependent $s^* \to VV$. 
This sub-amplitude is only one contribution to the full gauge-invariant $ss\to VV$ cross section,
and does not provide the dominant contribution in the hard limit $\hat{s}\gtrsim M_V^2 \gg m_s^2$.
Indeed, the cross section approaches a constant value at $\hat{s}\gtrsim M_V^2 $
because a vector with mass $M_V\ll\sqrt{\hat{s}} $ mediates a Coulomb-like long range force, leading to a soft/collinear enhancement
from the $t$-channel and $u$-channel diagrams.
We thereby write the result as
\beq\label{eq:ss2VV}
\hat\sigma(ss \to VV) = \frac{M_V^2}{4\pi w^4}\left[ R  \left(1+
\frac{3 M_V^2}{ 2\hat{s}}+12 \frac{M_V^4}{\hat s^2}\right)-
\frac{24 M_V^4  (\hat{s}-2 M_V^2) }{ \hat{s}^3}{\rm arctanh} \, R  
\right]\eeq
where $R=\sqrt{1-4 M_V^2/\hat{s}}$.
The factor in parenthesis approaches unity at large energy, 
and approaches $11 R/8$ around the threshold at $R\ll 1$.

Integrating over the parton density gives the production rate
\beq \label{eq:IV}\frac{N_{VV}}{A}= M_V^2 {\cal I}_V(\frac{M_V}{\gamma  m_s})+{\cal O}(\lambda g)^2
\eeq
where the dimension-less function ${\cal I}_V$ is plotted in fig.\fig{WallProductionRates}.

\smallskip

This can be compared to the off-shell results of~\cite{1211.5615,2403.03252}.
We confirm that the $s$ decay width is
\beq \Gamma(s \to VV) =  \frac{m_s^3}{32\pi w^2} \left(1 - 4\frac{M_V^2}{m_s^2} + 12\frac{M_V^4}{m_s^4} \right) \sqrt{1-4\frac{M_V^2}{m_s^2}}\eeq
and that its off-shell extension $\Pi(\hat s) $ at $\hat s\neq m_s^2$ is gauge dependent.
The enhancement at $\hat s\gg M_V^2$ claimed by~\cite{1211.5615},
$\Im\Pi \sim g^2 (3 M_V^2 - \hat s + \hat s^2/M_V^2)$, is a gauge artefact~\cite{2403.03252}.
The gauge dependence was tentatively bypassed in~\cite{2403.03252} relying on the Goldstone equivalence theorem
(that applies for $\hat{s}\gg M^2_V$), 
claiming a result similar to the scalar case, eq.\eq{Piscalar}:
\beq  \label{eq:Pivector}
\Im \Pi(\hat{s}) \sim  (2 g^4+\lambda^2)   \left[ \frac{w^2}{4\pi} + \frac{\hat{s}}{(4\pi)^3}\right]
  .\eeq
The first two body decay has a super-renormalizable suppression, avoided in the 3-body term.
In this case, the on-shell approach predicts a mildly larger $V$ production rate,
despite a smaller energy transfer.

\subsection{Graviton production}\label{sec:grav}
Finally, we consider production of hard gravitons $g$.
Graviton pair production arises from the covariantization of the scalar kinetic term,
\beq \hat\sigma(ss\to gg)=\frac{\hat{s}}{480\pi \bp^4}.\eeq
This model-independent effect is present only in the on-shell approach, but it's suppressed by 4 powers of the reduced Planck mass.
The covariantization of model-dependent interactions of $s$ allows single graviton production 
with rate suppressed by only two powers of $\bp$, so that
a fraction $\sim (\gamma m_s/\bp)^2$ of the total high-energy rate
is estimated to go into gravitons.

\section{Implications for dark matter}\label{sec:models}
We next study phenomenological implications of our revised rate of heavy particle production from ultra-relativistic bubble collisions.
We here start considering the possibility that some particle produced by wall collisions is a stable dark matter (DM) candidate, 
and estimate the resulting DM cosmological abundance.

We first derive the standard model-independent estimate for the dark matter yield sourced by bubble collisions.
We approximate the phase transition in terms of the usual dimensionless factors: 
$\alpha = \Delta V/\rho_{\rm rad}$ at nucleation controls the strength of the phase transition;
$\beta/H$ controls its inverse velocity;
$c_V = \Delta V/(2w)^4$ is related to the scalar quartic.
The average cosmological temperature when the phase transition ends is given by $\pi^2 g_* T^4_{\rm reh}/30 \approx \Delta V (1+1/\alpha)$
with $g_* =106.75$ degrees of freedom in the SM.
At this moment, the average cosmological abundance of DM particles produced in pairs from bubble collisions is
\beq n_{\rm DM} \approx \frac22 \frac{4\pi R_*^2}{4\pi R_*^3/3} \frac{N}{A}\approx \frac{3}{R_*}\frac{N}{A}\eeq
having approximated the bubbles as spheres with radius $R_* \approx (8\pi)^{1/3}/\beta$.
Next, if no other interaction happens, the ratio $Y_{\rm DM} = n_{\rm DM}/s$ remains constant, where $s =2\pi^2 g_s T^3/45$ is the entropy density
with $g_s\approx g_*$ at $T \gg \TeV$.
Then the present DM density is (see e.g.~\cite{2406.01705})
\beq 
\label{eq:OmegaDMY}
\Omega_{\rm DM} = \frac{\rho_{\rm DM}}{\rho_{\rm cr}} 
= \frac{688 \pi^3 \, T_0^3 \, Y_{\rm DM} \, M}{1485 \, M_{\rm Pl}^2 \, H_0^2}.
\eeq
Inserting numerical factors, the DM abundance becomes
\beq  \label{eq:OmegaDM}
\Omega_{\rm DM}h^2= 0.1 \frac{\beta}{H} \frac{M w}{(100\TeV)^2}\left[\frac{\alpha}{(1+\alpha) c_V}\right]^{1/4} \frac{1}{w^2}\frac{N}{A}
\eeq
where the dimension-less $(N/A)/w^2$ controls DM particle production in wall collisions, as computed in the earlier sections.
Cosmological measurements find $ \Omega_{\rm DM}h^2=0.1200\pm 0.0012$~\cite{2406.01705}.

\begin{figure}[t]
$$\includegraphics[width=0.9\textwidth]{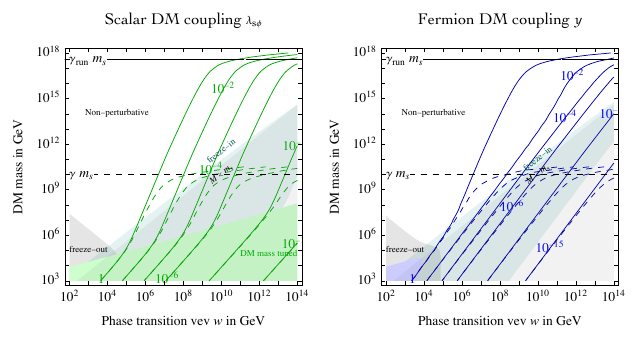}$$
\begin{center}
\caption{\em\label{fig:WallDMCouplings}\em Assuming the phase transition parameters of eq.\eq{params}, 
we show the contour values of the coupling $\lambda_{s\phi}$ of scalar (green, left) and 
$y$ of fermionic (blue) DM to the wall scalar needed to match the observed
DM abundance assuming that wall collisions and no other process contributes to DM production.
In the shaded region the DM is lighter than the contribution to its mass from the symmetry breaking.
Above the horizontal line the DM is too heavy to be produced in wall collisions, for the assumed value of the Lorentz factor $\gamma$.
}
\end{center}
\end{figure}

\smallskip

We present a numerical example assuming a phase transition with the same parameters as in~\cite{2403.03252}
\beq \label{eq:params}
c_V =0.1,\qquad \frac{\beta}{H}=10, \qquad\alpha=1,\qquad \ell_0=\frac{1}{2w},\qquad 
\gamma_{\rm run} m_s \approx 3.7~10^{17}\GeV\eeq
which corresponds to a sizeable $\lambda=8$.
Fig.\fig{WallDMCouplings}a  shows contours of the value of the coupling $\lambda_{s\phi}$ to a DM scalar $\phi$
needed to reproduce the DM abundance from bubble collisions, assuming
that $Y_{\rm DM}$  remains constant after production.
Fig.\fig{WallDMCouplings}b does the same for fermion DM.
In both cases we assumed the same Lagrangian interactions as in section~\ref{sec:cross},
$ -\lambda_{s\phi} s^2 \phi^2/4$ for scalar $\phi$ DM, and  $y s\bar{f} f/\sqrt{2}$ for fermion $f$ DM.
The upper horizontal line in our figure shows the  suppression at $M \gtrsim \gamma_{\rm run} m_s$, when DM is too heavy to be produced.
Assuming a lower $\gamma m_s$ lowers this maximal $M$, while mildly affecting the rest of the figure (dashed curves,
for $\gamma m_s =10^{10}\GeV$).

\smallskip

Fig.\fig{WallDMCouplings}  differs significantly from the analogous figures in~\cite{2403.03252}.
A key difference is that we find that the DM abundance cannot be reproduced in the upper-left region,
corresponding to $M \gg w$.
The reason is that the needed couplings become non-perturbative.
Indeed, for order unity couplings one can neglect $\lambda$ so that eq.\eq{OmegaDM}  gives the rough estimate
$ y^2, \lambda_{s\phi} \sim 100\TeV (M/w^3)^{1/2}$.
In the bottom-right region DM is lighter than the scalar driving the phase transition, so the DM production rate is large but
dominated outside the hard regime where the partonic approximation holds.
In the grey-shaded region labeled ``freeze-out'' dark matter re-thermalizes, while the teal region labeled ``freeze-in'' indicates where the freeze-in contribution is estimated to exceed that from bubble collisions.
These two processes have been estimated considering only the couplings to the walls scalars;
in general DM can have other interactions to SM particles, see~\cite{2406.01705} for a summary.
Even with our reduced rates, bubble collisions open up new dark matter parameter space 
where the dark matter mass is sufficiently above the reheating temperature to avoid later freeze-in and freeze-out.

\subsection{The wall Lorentz factor}\label{sec:gamma}
We next consider specific theories that allow to predict the parameters $\alpha,\beta,\gamma$
of the phase transition.
As usual, the space-time density of the bubble nucleation rate 
is computed in terms of the action $S_3$ of the usually dominant O(3)-invariant thermal bounce
as
\be\label{eq:Gamma_rate_S3T}
 \Gamma_{\rm nuc} \approx  T^{4}\left(\frac{S_{3}}{2\pi T}\right)^{3/2}\, e^{-S_3/T}.
\ee
Super-cooling is ended by bubble nucleations at the temperature $T_{\rm nuc}$ such that
$\Gamma_{\rm nuc}(T_{\rm nuc}) \approx H^{4}(T_{\rm nuc})$,
corresponding to $S_3/T\sim 100$.
Weak coupling generically gives a significant amount of super-cooling, $T_{\rm nuc}\ll \Delta V^{1/4}$.
The velocity of the phase transition
${\beta}/{H} = -{d \ln \Gamma_{\rm nuc}}/{d \ln T} |_{T=T_{\rm nuc}}$ 
often ranges one or two orders of magnitude around $\beta/H \sim 100$.

\smallskip

Bubble walls produced during nearly-vacuum first-order phase transitions can become ultra-relativistic and, 
in the absence of sufficiently strong friction, enter a runaway regime. 
The resulting Lorentz factor at collision $\gamma\gg1 $
controls both the maximal kinematic reach and the normalization of the wall parton luminosity. 
However, determining the Lorentz factor reached by the walls at collision remains an open problem, subject to significant theoretical uncertainties.

Neglecting friction, bubbles collide with the maximal Lorentz factor $\gamma_{\rm run} \sim R_*/R_0$ where $R_0 \approx 3\sigma/\Delta V \sim w^{-1}$, leading to a nearly Planckian energy
\begin{equation}\label{eq:gammarun}
    \gamma_{\rm run} \sim (8\pi)^{1/3}\sqrt{3}  \frac{w \bar{M}_{\rm Pl} }{\sqrt{\Delta V}} \frac{H}{\beta}\left(\frac{\alpha}{1+\alpha} \right)^{1/2} .
\end{equation}
In the presence of friction, the walls accelerate until the frictional pressure $\wp$ balances the 
accelerating force due to the vacuum energy difference $\Delta V$.
Some contributions to friction do not grow with $\gamma$, and are thereby equivalent to a reduced $\Delta V$. 
One example of this is plasma particles gaining mass inside the bubble.
One contribution gives a pressure that grows with $\gamma^p$ with some power $p$:
transition radiation emitted when particles enter the bubble~\cite{1703.08215}.
Vectors exhibit a peculiar radiation behaviour, that also leads to eq.\eq{ss2VV}. 
In theories where $s$ couples to a vector with gauge coupling $g$ and mass $M_V=gv$,
emission of one soft vector gives pressure $\wp \sim \gamma g^2 M_V T^3$~\cite{1703.08215}, 
limiting $\gamma \lesssim \Delta V  /g^2 m T_{\rm nuc}^3$.
This effect gets reduced down to $p=4/7$ in a saturation regime~\cite{1703.08215}.
Emission of multiple vectors was claimed to increase $p=2$~\cite{2005.10875,2007.10343}, but
under assumptions that have been criticised~\cite{2112.07686,2403.05615}; see~\cite{2305.02357} for a review.
Emission of scalars or fermions is not IR-enhanced and gives $\wp \propto \ln\gamma$. 

Given these uncertainties, we will conservatively show results for multiple values of $\gamma$.
Some phenomenological studies treat the Lorentz factor at collision as a free parameter.


\subsection{Theories of dark matter and first order phase transitions}\label{sec:2models}
A broad variety of speculative theories can give rise to a first-order phase transition,
and a broad variety of speculative theories contains particle dark matter candidates.

To maximise the predictivity we focus on classically scale-invariant
theories in which electroweak symmetry breaking is induced radiatively. 
The resulting Coleman-Weinberg potentials naturally tend to produce first-order phase transitions, 
making such models relevant for our discussion. 
Since the observed Higgs boson is too heavy for the minimal
Coleman-Weinberg realization, viable models typically extend the Standard Model
by an additional singlet scalar $s$, often called the dilaton. 
Further fields and interactions are then needed to generate the radiative symmetry breaking.
These extra fields can be stable and provide dark matter candidates.
In this minimal set-up all mass scales are generated dynamically, 
so the dark matter mass cannot be much heavier than the reheating temperature after the phase transition.
In the rest of this section we compute two specific simple models.

We anticipate the result:
in both cases we will find that bubble collisions can generate all observed DM,
despite that the on-shell approach predicts a smaller abundance than the off-shell approach.
However, in both models order unity factors imply that DM is not sufficiently heavier than the reheating temperature,
such that freeze-in generates a larger abundance, 
with freeze-out reducing the abundance in some regions of the parameter space.
This happens because in such models DM plays one extra role: 
dynamically inducing the first order phase transition that triggers the weak scale.
These models are intentionally minimal and somewhat self-limiting.

This discussion thereby implies that DM from bubble collisions can be phenomenologically significant in more general but
non-minimal models.
For example, one can  abandon scale invariance by adding an extra mass term for DM.
Alternatively, one can add a different massive DM component, without requiring that its couplings trigger dynamical symmetry breaking.
Along these lines one could consider models where a warped extra dimension is dual to some unspecified dynamics that breaks scale invariance, such as~\cite{1104.4791}.
Another possibility is dropping the connection with the weak scale.

These attempts risk hitting one common issue: if DM is too heavy and too strongly coupled,
its quantum corrections  affect the scalar potential and the phase transition. 
This issue can be circumvented by tuning the mass of the scalar involved in the phase transition.
In this context, one  can consider theories where the Higgs itself undergoes a first order phase transition,
 as the Higgs mass is anyhow unnaturally small, possibly due to anthropic selection.
DM can be added to such models along the lines of~\cite{1903.03616}.

\smallskip

To conclude this section, we present the computations in the two minimal models.

\begin{figure}[t]
\includegraphics[width=0.45\textwidth]{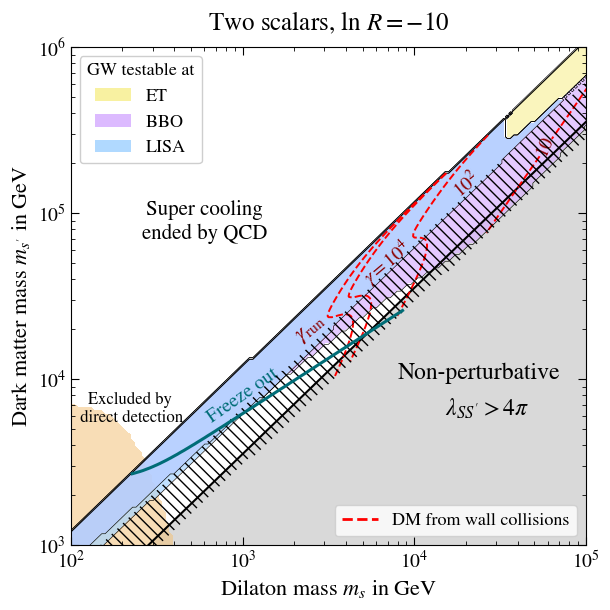}\qquad
\includegraphics[width=0.45\textwidth]{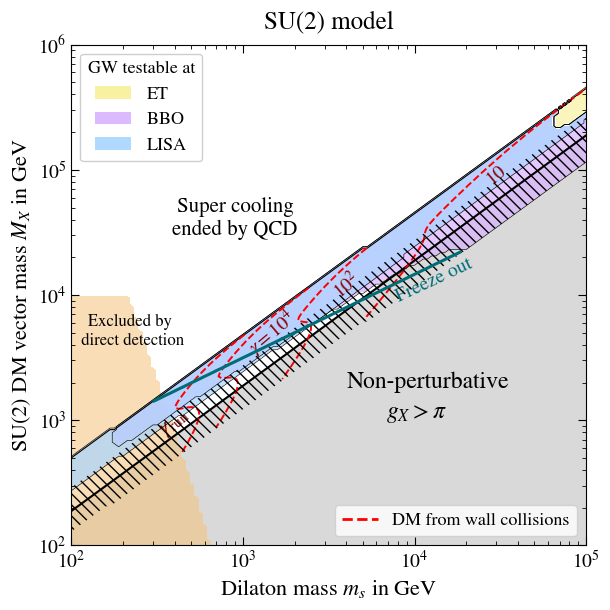}
\caption{\label{fig:ModelsDM}\it Parameter space of the two models of dynamical EW symmetry breaking
in section~\ref{sec:2models}, as function of the dilaton mass $m_s$ and of the DM mass.
The observed DM abundance is reproduced by thermal freeze-out along the darker green curve.
Wall collisions would reproduce the DM abundance along the red dashed curves
with the indicated Lorentz factor $\gamma$ (from 10 up to the maximal $\gamma_{\rm run}$).
Bubbles don't go runaway in the hatched regions.
However in these models the DM abundance gets later modified by freeze-in and freeze-out.
Gravitational waves from bubble collisions are detectable in the shaded regions, see~\cite{2409.04545}.}
 \end{figure}

\subsubsection*{Scalar DM}
A minimal model involves two extra scalars $s$ and $s'$ besides the Standard Model Higgs doublet $H$. 
Assuming that the  theory is separately invariant under $s\to - s$ and $s'\to - s'$, the symmetry  $\mathbb{Z}_2\otimes\mathbb{Z}'_2$
allows the scale invariant potential
\begin{equation}
\begin{split}
V_{\rm tree}
=
&\,V_\Lambda
+\lambda_H|H|^4
+\frac{\lambda_S}{4}s^4
+\frac{\lambda_{S'}}{4}s'^4
+\frac{\lambda_{HS}}{2}|H|^2s^2
+\frac{\lambda_{HS'}}{2}|H|^2s'^2
+\frac{\lambda_{SS'}}{4}s^2s'^2 .
\end{split}
\label{eq.Vtree}
\end{equation}
Following~\cite{2204.01744,2409.04545}, we can use as parameters of this model
the dilaton mass $m_s$, the DM mass $m_{s'}$ and one dimension-less
 parameter $R$ that controls which effective running coupling turns negative first,
 triggering electro-weak radiative symmetry breaking 
\be
\lambda_S^{\rm eff}(s) = \frac{{\beta}_{\lambda_S }}{2}   \ln \frac{s^2}{e^{1/2}w^2 },\qquad
\lambda_{HS}^{\rm eff}(s) = \frac{{\beta}_{\lambda_{HS} }}{2} \ln \frac{R s^2 }{w^2}.
\label{eq3}
\ee
Fig.\fig{ModelsDM}a recalls the phenomenology of the model in the $(m_s, m_{s'})$ plane (see e.g.~\cite{2204.01744,2306.17158}).
In the lower region $\lambda_{SS'} = (4\pi m_s)^2/m_{s'}^2$ is non perturbatively large.
In the upper region the phase transition is of second order, triggered by QCD~\cite{1704.04955}.
In the intermediate region the phase transition is of first order, triggered by bubble nucleation.
It is characterised by  $\alpha\gg 1$ and $\beta/H \sim 10 $ around the QCD boundary.
Bubbles go runaway if $\Delta V \gtrsim p_{\rm LO} \approx m_{s'}^2 T_{\rm nuc}^2/24$~\cite{0903.4099,2403.05615},
which results into $M_{s'}/T_{\rm nuc} \gtrsim 4\pi/\sqrt{3}$.

The red curves indicate where dark matter production from bubble collisions alone would reproduce the observed cosmological abundance.
We assumed an approximate tanh wall profile, and applied the on-shell formul\ae{} derived in the previous section.
We considered different values of the uncertain Lorentz factor $\gamma$, from $10$ up to the 
maximal possible $\gamma_{\rm run}$ of eq.\eq{gammarun}.
The dashed red curves have an $S$-like shape because the phase-transition parameter $\alpha$ is larger around the QCD border,
while $\beta/H$ is larger around the non-perturbative border.
We see that the contribution from bubble collisions can be significant.

However, in this model the reheating temperature $T_{\rm reh}$ is mildly above the freeze-out decoupling temperature $T_{\rm dec}\approx m_{s'}/25$.
This proximity is not an accident: it happens because $\Delta V$ is radiatively induced by the quantum fluctuations of DM itself.
So freeze-in and freeze-out largely correct the DM abundance after the phase transition.
The cosmological DM abundance is thereby reproduced as thermal freeze-out along the green curve, as in previous computations~\cite{1805.01473,2204.01744,2306.17158}.
In the regions above this curve, the DM abundance is too large.
Similar results arise for other values of the $\ln R$ parameter.

\subsubsection*{$\SU(2)_X$ vector DM}
As a second model, we consider adding to the SM  a complex scalar doublet $S$  under an extra dark 
$\SU(2)_X$ gauge group with gauge coupling $g_X$~\cite{1805.01473}.
The scale-invariant potential is 
\beq  V_{\rm tree} = V_\Lambda + \lambda_H |H|^4 + \lambda_S |S|^4 + \lambda_{HS} |H|^2| S|^2 .\label{eq:V}\eeq
The radial component of $S$ is the scalar $s$.
The $\SU(2)_X$ vectors acquire mass $M_X = g_X w/2$ and are stable DM candidates
without imposing any ad-hoc $\mathbb{Z}_2$ symmetry~\cite{0811.0172}.
As a result $XX\leftrightarrow X s$ semi-annihilations are present, in addition to DM annihilations.
Their thermal abundance matches the cosmological DM abundance when 
\beq  \sigma_{\rm ann} + \frac{1}{2}\sigma_{\rm semi-ann}=
\frac{19 g_X^4}{3456\pi M_X^2}\eeq
equals $\approx 1/(23\TeV)^2$.
As in the previous model, we again use as free parameters the dilaton mass $m_s$ and the DM mass $M_X$.
Bubbles don't go runaway if $\Delta V \lesssim  p_{\rm LO} \approx 3 M_X^2 T_{\rm nuc}^2/8$~\cite{0903.4099,2403.05615}, which gives
$M_X/T_{\rm nuc} \lesssim 4\pi/\sqrt{3}$: this region is hatched in fig.\fig{ModelsDM}.
In this model bubble collisions produce vector DM with the enhanced cross section of eq.\eq{ss2VV}.
Fig.\fig{ModelsDM}b shows that this can match the cosmological DM abundance.
However, the reheating temperature after the phase transition is again too high,
so that freeze-out and freeze-in restore the dark matter abundance to the results obtained in previous studies, 
leaving it unaffected by bubble collisions.


\section{Implications for leptogenesis}\label{sec:leptog}
Thermal leptogenesis~\cite{Fukugita:1986hr} can reproduce the observed baryon asymmetry
\beq n_B/n_\gamma|_{\rm exp} \approx 6.1~10^{-10},\eeq 
provided that the mass of the lightest right-handed neutrino $N$ is $M_1 \gtrsim 10^{9}\GeV$
and that the reheating temperature after inflation is above $M_1$, see e.g.~\cite{hep-ph/0310123}.
We consider the possibility that a first order phase transition occurred.
Two main questions are important, in two different regimes:
\begin{enumerate}
\item If the inflationary reheating temperature is below $M_1$ thermal leptogenesis fails.
Can leptogenesis nevertheless happen provided that right-handed neutrinos are produced by
runaway bubbles that collide with large enough Lorentz factors, $\gamma m_s  \gtrsim M_1$?

\item If the inflationary reheating temperature is above $M_1$,
successful thermal leptogenesis can be ruined by the super-cooling associated
with the first order phase transition, if it significantly dilutes 
the previously generated baryon asymmetry.
Can leptogenesis be re-done by bubble collisions?

\end{enumerate}
According to the off-shell approach both issues have a positive answer, as 
the  $N$ abundance produced by bubble collisions is easily enough for leptogenesis~\cite{2407.16747,2601.02458}.
However the off-shell approach can largely over-estimate the $N$ abundance.
So we reconsider both issues in light of our revised rate of particle production from bubble collisions.

\smallskip

Different models are possible, and we focus on the simpler possibility.
We assume that the right-handed neutrino $N$ has a Yukawa coupling
to the scalar $s$ involved in a first order phase transition~\cite{2601.02458},
\beq\Lag   \supset \left[(\bar{M}_1+ \frac{y\, s}{\sqrt{2}}) \frac{N^2}{2} +y_N N L H  +\hbox{h.c.}\right].\eeq
In a full theory $s$ can be the radial component of a complex scalar $S$ that spontaneously breaks a U(1)$_{B-L}$
global or gauged symmetry, preventing an extra mass term for $N$.
The $N$ abundance produced by bubble collisions is 
\beq Y_N = \frac{n_{N}}{s} \approx  \frac{1}{s} \frac{3}{R_*}\frac{N}{A} \approx 0.45 \frac{w}{M_{\rm Pl}}\frac{\beta}{H}
\bigg[\frac{\alpha}{(1+\alpha)c_V}\bigg]^{1/4} \frac{1}{w^2}\frac{N}{A}
\eeq
where $s$ is the entropy density.
According to the on-shell result of eq.\eq{If}
the number of $N$ produced per unit area is $(N/A)/w^2 \sim y^4 w^2/M^2_1 + \cdots$, 
which can reach order unity.
The corresponding off-shell over-estimate is $(N/A)/w^2 \sim y^2$.

\begin{figure}[t]
$$\includegraphics[width=0.45\textwidth]{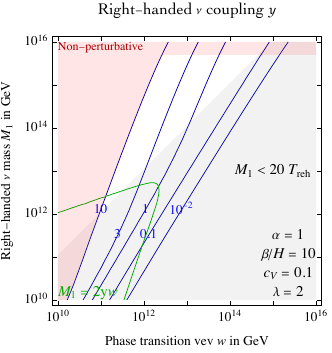}\qquad
\includegraphics[width=0.45\textwidth]{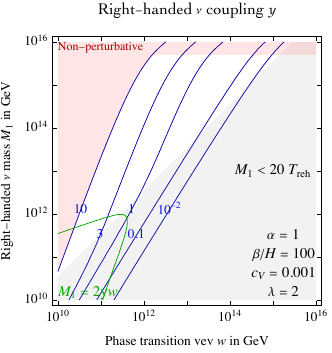}$$
\caption{\label{fig:Leptog}\it 
We show contour values of the right-handed neutrino coupling
$y$ to the wall scalar needed to match the observed baryon asymmetry.
The left panel assumes the phase transition parameters of eq.\eq{params},
the right panel assumes the indicated more favourable values.
The grey region approximates where $M_1$ is so light to be dominantly produced by freeze-in
via scattering of SM particles.
The red region indicates where either $y$ or a right-handed neutrino coupling to leptons
is non-perturbatively large.
Along the green curve $M_1$ entirely arises from the Yukawa coupling to $s$.
}
 \end{figure}

Ignoring flavour issues, the baryon asymmetry produced from $N$ decays is usually approximated as
\beq Y_B=-\frac{28}{79} Y_N \epsilon_1 \eta,\qquad\hbox{where}\qquad
\epsilon_1 = \frac{3}{16\pi} \frac{M_1\Im \tilde m}{v_{\rm SM}^2}
\eeq
is the CP-violating asymmetry in $N\to LH,\bar L H^*$ decays, $v_{\rm SM}=174\GeV$ and $\tilde{m}$ is the contribution to neutrino masses not mediated by $N$.
This means that a large CP asymmetry needs a large $M_1$.\footnote{The CP asymmetry gets resonantly enhanced if two right-handed neutrinos are quasi-degenerate,
but in this case thermal leptogenesis too can work at low reheating temperature, without needing bubble collisions.}
Furthermore,  $\eta\le 1$ is a dynamical efficiency factor.
It plausibly becomes $\eta \simeq 1$ for bubble collisions, as $N$
decays out of equilibrium.\footnote{Thermal leptogenesis is
problematic in type II~\cite{hep-ph/0510008} and type III~\cite{0806.1630} see-saw models because neutrino masses are
mediated by particles with weak interaction, that remain close to thermal equilibrium.
In these cases production from ultra-relativistic bubble collisions would largely enhance the efficiency.}
Using $s \approx 7.04 n_\gamma$ gives
\beq \frac{n_B}{n_B|_{\rm exp}}\approx \frac{\Im\tilde m}{0.05\eV} \frac{\beta}{H} \frac{w M_1}{(2.6~10^{12}\GeV)^2}
\bigg[\frac{\alpha}{(1+\alpha)c_V}\bigg]^{1/4}  
\frac{1}{w^2}\frac{N}{A}.
\eeq
If the off-shell result $N/Aw^2 \sim y^2$ was correct, the baryon asymmetry could have been reproduced 
at fixed $y\sim1$ for a fixed value of the product $wM_1$. 
Taking $M_1$ sufficiently large would then have allowed access to the phenomenologically interesting region of small $w$, 
in which bubble-wall collisions could enable leptogenesis even for a low post-inflationary reheating temperature, $T_{\rm RH}\gtrsim c_V^{1/4} w$.
The on-shell calculation, however, introduces an additional suppression factor $(w/M_1)^2$ relative to the off-shell estimate. Consequently, $w$ must also be large, roughly $w\gtrsim10^{10}\GeV$. A low reheating temperature can then be obtained only for phase transitions with $c_V\ll1$, which requires a significant tuning, since right-handed neutrinos themselves contribute to the vacuum energy.
We therefore conclude that leptogenesis remains viable only within a restricted region of parameter space: the answer to question 1 is affirmative, whereas the answer to question 2 is negative.

This is illustrated in fig.\fig{Leptog}. 
The left panel assumes the phase-transition parameters of eq.\eq{params}, whereas the right panel adopts more favourable values of $\beta$ and $c_V$. In both panels we take $\Im\tilde{m}=0.05\eV$, a representative value compatible with current knowledge of neutrino masses.
Leptogenesis from $N_1$ particles produced in bubble-wall collisions is successful in the white region, where the Yukawa coupling $y$ is sufficiently large. This region includes the green curve corresponding to $\bar{M}_1=0$, for which the right-handed neutrino mass arises entirely from its coupling to $s$, as naturally occurs in ${\rm U}(1)_{B-L}$ models. 
By contrast, the off-shell estimate would have incorrectly identified the entire upper-left region of both panels as viable.

\smallskip

In conclusion, minimal leptogenesis from bubble-wall collisions requires a phase transition at a relatively large scale $w$, and hence a correspondingly high reheating temperature $T_{\rm reh}$. 
The gravitational-wave signal generated by the transition is therefore expected to peak around the frequency
$f \sim {\rm Hz} (\beta/H) (T_{\rm reh}/10^7\GeV)$.

\section{Implications for gravitational waves}\label{sec:GW}
In section~\ref{sec:GW1} we reconsider gravitational waves sourced by the motion of the particles produced during bubble collisions.
In section~\ref{sec:GW2} we consider gravitons directly produced during bubble collisions.

\subsection{Gravitational waves from particle production}\label{sec:GW1}
If bubble collisions convert a sizeable amount
of wall energy into energetic particles, the subsequent free streaming of these
particles provides an additional source of gravitational waves, with a
characteristic spectrum~\cite{2412.17912,2601.02458}. The energy carried by the
produced particles per unit wall area $A$ can be written as
\beq
 \frac{E}{A} = \med{E}\,\frac{N}{A}\, .
\eeq
In the off-shell estimate the average particle energy is dominated by the
ultraviolet end of the wall spectrum,
$ \med{E}\sim \gamma m_s $,
so that the scattered energy is enhanced by the wall Lorentz factor. As a result,
a fraction
$  \kappa_{\rm sc}\equiv
 {E}/{E_{\rm wall}} \sim g^2$
of the total wall energy is converted into particles, where $g$ denotes a
representative dimension-less coupling, that can be of order unity.

This claim needs to be reconsidered, given that the off-shell results largely over-estimate the amount of hard scattering.
In the on-shell approach a typical cross section is $\hat{\sigma}\sim  g^4/\hat{s}$.
The convolution with the wall parton luminosity is then dominated by the infrared end of the spectrum, giving $\med{E} \sim m_s$.
Consequently the fraction of wall energy converted
into scattered particles is only $ \kappa_{\rm sc}\sim g^4/\gamma$ up to possible saturation effects.
Since the gravitational-wave amplitude is
linear in the transverse-traceless stress tensor, while the gravitational-wave
energy density is quadratic in it, the corresponding contribution to
$\Omega_{\rm GW}$ scales as $\kappa_{\rm sc}^2$.
So the off-shell approach over-estimated  $\Omega_{\rm GW}$ by a large $\gamma^2$ factor. 
In other words, the large effect claimed by the off-shell picture came from assuming order-one conversion of wall energy into hard scattered particles. Once hard scattering is treated microscopically, this large effect disappears.

\smallskip

The situation might seem different for vector production, whose cross section in
eq.\eq{ss2VV} is approximately constant at large energy. This enhancement,
however, is dominated by soft configurations with small $|\hat{t}|$ or $|\hat{u}|$: 
the scattered particles mostly keep moving along the initial wall directions, while
changing their identity from $s$ to $V$. 
The energy carried by the
final vectors is of order $\sqrt{\hat s}$, but the energy relevant for
deflecting the wall stress tensor is much smaller.
A transport-weighted estimate of the energy removed from the original wall
directions is
\beq  \frac{E}{A} \approx \int d\hat{s}  \frac{d{\cal L}_{ss}}{d\hat{s}} \int d\hat{t}  \,\frac{d\hat\sigma}{d\hat{t}} \,   \frac{\min (-{\hat{t}} , -{\hat{u}} )}{\sqrt{\hat{s}}}.\eeq
As a result the scattered energy is again suppressed by $1/\gamma$ compared to the off-shell claim.
Nevertheless, the large rate for the identity-changing collinear process $s\to V$ can
affect the subsequent collective evolution of the walls and plasma, and thereby
modify the usual gravitational-wave source. This is a different effect from
direct gravitational radiation sourced by hard, large-angle scattered particles.

In conclusion, the additional $1/\gamma^2$ suppression of the scattered particle contribution to
$\Omega_{\rm GW}$ renders it of little phenomenological interest for $\gamma\gg1 $.
The suppression gets smaller if walls are only mildly non-relativistic, $\gamma \sim {\rm few}$,
but this needs a large friction 
so that wall particle collisions anyhow involve a small fraction of the universe energy,
$ \kappa \equiv  3 (E/A)/2 R_* \Delta V \ll \kappa_{\rm sc}$.

\subsection{Direct graviton production}\label{sec:GW2}
Following section~\ref{sec:grav}, we consider gravitons singly produced from bubble collisions as
$ss \to g p$ where $g$ is a graviton and $p$ is some massless particle. 
The total energy  acquired by gravitons $g$ in the center of mass frame is
\beq \frac{E_g}{A}
= \int d\hat s   \frac{d{\cal L}_{ss}}{d\hat s} \frac{\sqrt{\hat s}}{2}
\hat\sigma (\hat s) = \med{E_g} \frac{N_g}{A}, 
\eeq
with $\med{E_g} \sim \gamma m_s$, as gravitational interactions are enhanced by high energy.
The produced graviton density  at  $T\approx T_{\rm nuc}$ is
\beq \rho_{g} \approx  \frac{3}{2R_*}\frac{\med{E_g}}{A},\qquad
\left. \frac{\rho_g}{\rho_{\rm SM}}\right|_{\rm production} \approx  
0.37  \frac{\med{E_g}}{M_{\rm Pl}} \bigg[\frac{\alpha}{(1+\alpha)c_V}\bigg]^{1/2}
\frac{\beta}{H} \frac{1}{w^2}\frac{N_g}{A}
\eeq
where $M_{\rm Pl} = \sqrt{8\pi}\bp = 1.2~10^{19}\GeV$ is the Planck mass.
This is converted into the total $\Omega_{g}$ today as
\beq\label{eq:Omegag}
\Omega_{g} \equiv \frac{\rho_{g}(T_0)}{\rho_{\rm cr}} =
3.6~10^{-5}  \left.\frac{\rho_{g}}{\rho_{\rm SM}}\right|_{\rm production} .
\eeq
A typical rate $(1/w^2)(N_g/A) \sim  ( g w/M_{\rm Pl})^2$ induced by a dimension-less coupling $g$
leads to
\beq \label{eq:Omegag}
\Omega_g  \sim
3~10^{-5} \frac{\beta}{H} \frac{g^2 \med{E_g} w^2}{M_{\rm Pl}^3}  \bigg[\frac{\alpha}{(1+\alpha)c_V}\bigg]^{1/2}
.\eeq
A mildly sub-Planckian $\med{E_g}$ can saturate the CMB and BBN bound $\Omega_g \lesssim 2.5~10^{-6}$
on an extra relativistic species~\cite{Planck:2018vyg}
only if $w$ and $\beta/H$ are large and $c_V\ll 1$.

\smallskip

The frequency spectrum of the produced gravitons is peaked around some average $\med{f_0}$.
The  $\gamma\gg1$ enhancement allows to obtain a large $\Omega_g$
but also enhances the frequency of these gravitons today
to a range where gravitational wave detection is difficult~\cite{2412.17897}:
\beq
\med{f_0}=\frac{\med{E_{g}} }{2\pi} \frac{T_0}{T_{\rm reh}}
\left(\frac{g_{s0}}{g_{s}}\right)^{1/3} \sim
10^{10}\,{\rm Hz}\,
\frac{\gamma m_s}{T_{\rm reh}},\qquad T_{\rm reh}
=
\left[
\frac{30}{\pi^2 g_*}
\frac{1+\alpha}{\alpha}
\Delta V
\right]^{1/4}.
\eeq



\section{Conclusions}\label{sec:concl}
During first order phase transitions bubbles form and expand, 
possibly reaching large Lorentz factor $\gamma\gg 1$ when they finally collide.
Heavy particle production was computed from the imaginary part of the effective action of the colliding bubble field configuration.
This was approximated in terms of the off-shell decay width $\Im \Pi(\hat{s})$ of the scalar $s$ driving the phase transition.
The relevant field configuration was then evaluated under the assumption of a perfectly elastic collision.
As a result, this `off shell' formalism predicts that bubble collisions send all their quanta off-shell up to $\gamma m_s$~\citeall.
When vector production was considered, the result was found to be gauge dependent,
and a tentative workaround  
 was presented~\cite{2403.03252}.

\smallskip

We showed that this formalism has a more general problem, present also when scalars or fermions are produced.
Its result depends on the parameterisation chosen in field space, such that a non-vanishing rate can also arise in free theories.
The difficulty is not with the exact effective action itself, 
but with evaluating it over an approximate collision background,
and with approximating in terms of the propagator $\Im \Pi(\hat{s})$.
Off-shell quantities such as $\Im \Pi(\hat{s})$ away from the pole are in general unphysical.
We clarified that this formalism provides correct results when the emitting particle $s$ has negligible self-interactions.
Bubble collisions arise in the opposite limit, when the self-interaction of the wall scalar has non-perturbative classical effects.
As a result, formal prescriptions intended to repair the method are of little practical use.

\smallskip

We proposed a different approximation for heavy particle production from bubble collisions, entirely based on on-shell dynamics.
The key observation is that wall collisions simplify when the Lorentz factor is ultra-relativistic, $\gamma\gg 1$.
The walls are relativistically contracted so that the potential force has no time to act.
To leading order, the two walls therefore pass through one another according to the free-passage approximation, 
with corrections suppressed parametrically by $1/\gamma^2$.
We stress that momenta of order $\gamma m_s$ in a chosen frame are not equivalent to time-like invariant masses of that order.
The Fourier transform of the smooth collision falls as $1/\hat s^2$, 
while an elastic collision gives a qualitatively misleading $1/\hat s$ behaviour.
The interaction is suppressed at large $\hat{s}$,
and can be approximated taking into account that different parts of the wall do not have time to interact.
Particle production can then be approximated as an incoherent sum of on-shell scatterings among the quanta in the wall.
Later coherent evolution is important at times of order $1/m_s$, and 
it does not restore the unsuppressed hard time-like tail in the  collision factor.

The resulting rate is parametrically lower than in the off-shell approximation, because
cross sections for high-energy $E$  exchange scale as $\hat\sigma\propto 1/E^2$.
This restores the standard hard-scattering behaviour that follows from
a basic property of quantum mechanics: large energy means small length.

\medskip

We next reconsidered phenomenological implications of particle production from bubble collisions  in light of the reduced rate.
We computed the parton luminosity of a basic tanh wall profile and
the generic cross sections for production of scalars, fermions, vectors.
We applied these ingredients to production of dark matter particles, finding that the cosmological DM abundance can be produced
from bubble collisions, despite the reduced rate compared to the incorrect off-shell earlier results.
We found that bubble collisions can produce enough right-handed neutrinos for leptogenesis,
but only in a restricted region of the parameter space that needs a reheating temperature as high as that required by thermal leptogenesis.
We found that gravitational wave production from the motion of particles scattered during wall collisions becomes a small effect,
suppressed by $\gamma$.
We considered production from bubble collisions of hard gravitons that today have
high frequencies $\sim \gamma T_0$.

\smallskip

In conclusion, our approach is analogous to the partonic approach used to compute hard processes in ultra-relativistic hadron collisions,
and has a similar applicability.
One might wonder whether our analysis overlooks some collective quantum effect
that might recover the much larger rates claimed by the off-shell approximation.
If such a mechanism were found, besides its cosmological implications, 
it could be applied to build colliders that probe high energy $E$ without needing a correspondingly high integrated luminosity,
${\cal L} \propto E^2$.

\footnotesize

\paragraph{Acknowledgments}
A.G.\ acknowledges the support from the Royal Society, UK, Funding Reference: NIF\ R1\ 253963 and Marek Lewicki for clarification.
A.S.\ thanks Claude, ChatGPT, Gian Giudice for clarifications, and
Tsuji Tenta for noticing an error in eq.\eq{Omegag}.

\end{document}